%% file: main.tex
\newcommand{\mode}{0}  % 2 for Journal Computational Physics, 1 for Physical Review B, 0 for JCTC
\ifcase\mode 0
  \documentclass[journal=jctcce,manuscript=article]{achemso}
\or 1
  \documentclass[aps,amsmath,amssymb,reprint]{revtex4-1}
\or 2
   \documentclass[10pt]{elsarticle}
\fi
\usepackage{comment}
\usepackage{braket}
\usepackage{subfigure}
\usepackage{graphicx}
\usepackage{amsmath}
\usepackage{url}
\usepackage{color}
\usepackage{verbatim}
\usepackage{gensymb}
\usepackage[utf8]{inputenc}
\usepackage{multirow}
\usepackage[english]{babel}
\usepackage{amsfonts}
\usepackage{amssymb}
\usepackage{enumerate}
\usepackage{sidecap}
\usepackage{bbold}
\usepackage[normalem]{ulem}
\usepackage{mathtools}
\usepackage[usenames, dvipsnames]{xcolor}
\usepackage{breqn}
\ifnum\mode>0
 \usepackage{natbib}
\fi
\usepackage[colorlinks=true,plainpages=true,citecolor=blue,linkcolor=blue]{hyperref}
\newcommand{\vcoul}{v_{\mathrm{c}}}
\newcommand{\EHartree}{\mathrm{E}_{\mathrm{H}}}
\newcommand{\Exc}{\mathrm{E}_{\mathrm{xc}}}

\newcommand{\fxc}{f_{\mathrm{xc}}}
\newcommand{\fxcb}{\bar{f}_{\mathrm{xc}}}
\newcommand{\fxcqp}{f_{\mathrm{xc}}^{\mathrm{QP}}}
\newcommand{\fxcks}{f_{\mathrm{xc}}^{\mathrm{KS}}}

\newcommand{\chioqp}{\chi_0^{\mathrm{QP}}}
\newcommand{\chioks}{\chi_0^{\mathrm{KS}}}
\newcommand{\chiomb}{\chi_0^{\mathrm{MB}}}
\newcommand{\chiad}{\chi_{\mathrm{adiab}}}
\newcommand{\chiksad}{\chi_{\mathrm{adiab}}^{\mathrm{KS}}}
\newcommand{\chiqpad}{\chi_{\mathrm{adiab}}^{\mathrm{QP}}}
\newcommand{\Woqp}{W_0^{\mathrm{QP}}}
\newcommand{\Woks}{W_0^{\mathrm{KS}}}
\newcommand{\Woqpij}{W_{0,ij}^{\mathrm{QP}}}
\newcommand{\Woksij}{W_{0,ij}^{\mathrm{KS}}}
\newcommand{\Prpaqp}{P_{\mathrm{RPA}}^{\mathrm{QP}}}
\newcommand{\Prpaqpijs}{P_{\mathrm{RPA},ij\sigma}^{\mathrm{QP}}}
\newcommand{\Prpaks}{P_{\mathrm{RPA}}^{\mathrm{KS}}}
\newcommand{\Prpaksijs}{P_{\mathrm{RPA},ij\sigma}^{\mathrm{KS}}}
\newcommand{\Gqp}{G^{\mathrm{QP}}}
\newcommand{\Gks}{G^{\mathrm{KS}}}
\newcommand{\Gd}{G^{\mathrm{D}}}
\newcommand{\Gout}{G^{\mathrm{out}}}
\newcommand{\Gin}{G^{\mathrm{in}}}
\newcommand{\Sigmaxc}{\Sigma_{\mathrm{xc}}}
\newcommand{\Sigmaxcijs}{\Sigma_{\mathrm{xc},ij\sigma}}
\newcommand{\Sigmaxcb}{\bar \Sigma_{\mathrm{xc}}}
\newcommand{\Ex}{\mathrm{E}_{\mathrm{x}}}
\newcommand{\Exb}{\bar{\mathrm{E}}_{\mathrm{x}}}
\newcommand{\Ec}{\mathrm{E}_{\mathrm{c}}}
\newcommand{\Ek}{\mathrm{E}_{\mathrm{k}}}
\newcommand{\vext}{v_{\mathrm{ext}}}
\newcommand{\vH}{v_{\mathrm{H}}}
\newcommand{\Ecb}{\bar{\mathrm{E}}_{\mathrm{c}}}
\newcommand{\Wbt}{\bar{\tilde{W}}}
\newcommand{\Eint}{\mathrm{E}^{\mathrm{inter}}}
\newcommand{\consistent}{consistent}

\newcommand{\appendixinfo}{Supporting Information}
\renewcommand{\Im}{\mathrm{Im}}
\newcommand{\lsi}{LSI, CNRS, CEA/DRF/IRAMIS, \'Ecole Polytechnique, Institut Polytechnique de Paris, F-91120 Palaiseau, France}
\newcommand{\etsf}{European Theoretical Spectroscopy Facility (ETSF)}
\newcommand{\GF}{GF}

\ifnum\mode=0  % this is JCTC case, where title,authors, aff. are in preamble
  \author{Abdallah El-Sahili}
  \affiliation{\lsi}
  \alsoaffiliation{\etsf}
  \email{abdallah.el-sahili@polytechnique.edu}
  \author{Francesco Sottile}%
  \affiliation{\lsi}
  \alsoaffiliation{\etsf}
  \author{Lucia Reining}
  \affiliation{\lsi}
  \alsoaffiliation{\etsf}
\title{Total energy beyond $GW$: exact results and guidelines for approximations}
\fi
\begin{document}

\ifcase\mode 2
 %\title{Total energy beyond $GW$: exact exchange-correlation energy via non-exact self-energy OR Going beyond the GW approximation using time-dependent density functional theory: a users' guide OR Combing Green's functions and time-dependent density functional theory for total energies and spectra OR Approximate self-energies, in principle exact results and guidelines for a use in practice (THE LATTER IS IS MY (LR) FAVORITE TITLE) OR Approximate self-energies, exact results and guidelines for approximations (FS)  }
 \title{Total energy beyond $GW$: Approximate self-energies, in principle exact results and guidelines for a use in practice (LUCIA)}
% \title{Total energy beyond $GW$: Approximate self-energies, exact results and guidelines for approximations (FS)  }
 \author[1,2]{Abdallah El-Sahili\corref{cor1}}
 \ead{abdallah.el-sahili@polytechnique.edu}
 \author[1,2]{Francesco Sottile}
 \author[1,2]{Lucia Reining}
 \affiliation[1]{organization={LSI, CNRS,  CEA/DRF/IRAMIS, Ecole Polytechnique, Institut Polytechnique de Paris},
                 postcode={F-91120},
                 city={Palaiseau},
                 country={France}}
 \affiliation[2]{organization={European Theoretical Spectroscopy Facility (ETSF)}}
 \cortext[cor1]{Corresponding author}
\or 1
  \title{Total energy beyond $GW$: exact exchange-correlation energy via non-exact self-energy OR Going beyond the GW approximation using time-dependent density functional theory: a users' guide OR Combing Green's functions and time-dependent density functional theory for total energies and spectra OR Approximate self-energies, in principle exact results and guidelines for a use in practice (THE LATTER IS IS MY (LR) FAVORITE TITLE) OR Approximate self-energies, exact results and guidelines for approximations (FS)  }
  \author{Abdallah El-Sahili}
  \affiliation{\etsf}
  \affiliation{\lsi}
  \email{abdallah.el-sahili@polytechnique.edu}
  \author{Francesco Sottile}%
  \affiliation{\lsi}
  \affiliation{\etsf}
  \author{Lucia Reining}
  \affiliation{\lsi}
  \affiliation{\etsf}
\fi
\date{\today}

\input{abstract}
\maketitle
\begin{tocentry}
\includegraphics[width=\textwidth]{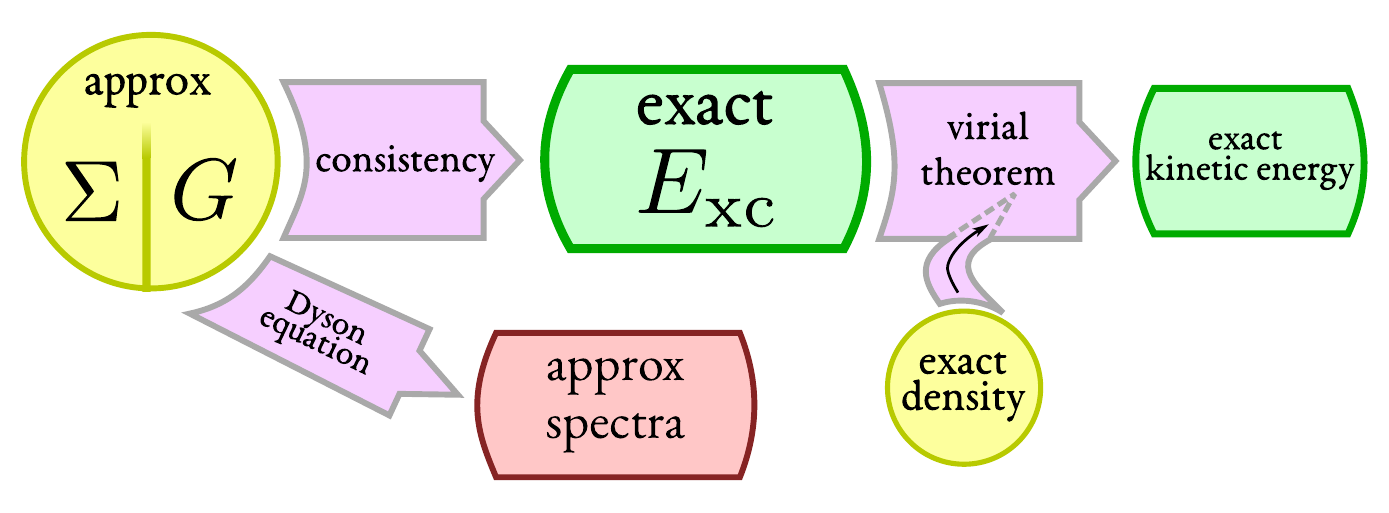}
The Dyson equation with an approximate self-energy leads to an approximate Green's            
function and to approximate spectra.
However, the same approximate self-energy yields the exact                 
exchange-correlation energy, provided the latter is evaluated following a consistent             
prescription.
\end{tocentry}

\input{introduction}

\input{theoretical_background_section}

\input{theoretical_development_section}

\input{illustrations_new}

\input{conclusion}
\input{appendices}

\bibliography{nlvc,biblio}

\end{document}

%% file: abstract.tex
\begin{abstract}
The total energy and electron addition and removal spectra can in principle be obtained exactly from the one-body Green's function. In practice, the Green's function is obtained from an approximate self-energy. In the framework of many-body perturbation theory, we derive different expressions that are based on an approximate self-energy, but that yield nevertheless in principle the exact exchange-correlation contribution to the total energy for any interaction strength. Response functions play a crucial role, which explains why, for example, ingredients from time-dependent density functional theory can be used to build these approximate self-energies. We show that the key requirement for obtaining exact results is the consistent combination of ingredients. Also when further approximations are made, as it is necessary in practice, this consistency remains the key to obtain good results. All findings are illustrated using the exactly solvable symmetric Hubbard dimer.

\end{abstract}

%% file: introduction.tex
\section{Introduction}
Many important properties of materials are linked to observables that can be expressed in principle as expectation values in the many-body ground state or in thermal equilibrium. In practice, the use of many-body wavefunctions is often avoided by rather describing the observables as functionals of more compact quantities, i.e., quantities that depend on less arguments, such as the density  \cite{Hohenberg1964}, one-body reduced density matrix \cite{Gilbert1975,Donnelly2008,Levy1979}, or one- or two-body Green's functions \cite{Martin2016}. This represents a trade-off: often, one does not know the exact functional for an observable in terms of these quantities, and approximations have to be designed. One important example is the total energy: it can be straightforwardly formulated in terms of the one-body Green's function (\GF) \cite{Galitskii1950}, whereas no exact explicit expression in terms of the density is known. The same holds for electron addition and removal spectral functions. Excitation spectra involving neutral excitations in linear response, instead, are easily expressed in terms of a two-body Green's function, but not in terms of the ground state density or the one-body Green's function \cite{Martin2016}. Even when the expressions are known, one faces another problem: while the use of the compact quantities carries the promise of reduced computational load, they are themselves only known explicitly as expectation values involving many-body wavefunctions. Therefore, nothing is gained, unless one finds ways to calculate them in a different way, which may be in principle exact, and in practice, require approximations. Typically, the density is obtained from the Kohn-Sham equations \cite{Kohn1965} with an approximate exchange-correlation (xc) potential, and the \GF , from a Dyson equation with an approximate xc self-energy $\Sigma_{\rm xc}$ \cite{mahan-mpp}. It is therefore not always obvious which framework (Density Functional Theory (DFT), Green's Function Functional Theory, etc.) is the best choice to access a given observable.

This holds in particular for the total ground state energy $\text{E}_0$. While the Galitskii-Migdal formula \cite{Galitskii1950} or functional expressions such as the Luttinger-Ward \cite{Luttinger1960} or Klein \cite{Klein1961} functionals yield an in principal exact and, in the latter two cases, even variational, form in terms of the Green's function $G$, the need for approximations to the \GF\ itself strongly impacts the quality of the results. Together with a computational load far heavier than that of the most widely used density functionals, this explains why the vast majority of total energy calculations is performed using DFT, not Green's functions. Still, research concerning total energy calculations using Green's functions is active and important \cite{Holm1999,Garcia2001,Caruso2012,Bruneval2021b}. Besides the - important - fact that in principle expressions for $\text{E}_0$ as functional of $G$ and/or $\Sigma_{\rm xc}$ are known, the Green's functions framework benefits from the existence of powerful approximations. In particular, Many-Body Perturbation Theory (MBPT) \cite{mahan-mpp} suggests a way to expand the self-energy in diagrams that carry physical meaning and that are therefore helpful to describe phenomena such as the van der Waals dispersion interaction \cite{Garcia2002}. For situations that show only weak to moderate interaction effects, MBPT is often considered to be a systematic way to proceed, although in practice renormalizations, such as screening of the Coulomb interaction, are needed. In particular, even the lowest order of an expansion of the self-energy in terms of the screened Coulomb interaction $W$, which is the widely used $GW$ approximation \cite{Hedin1965}, has been very successful for the calculation of the quasi-particle (QP) part of electron addition and removal spectra in finite and extended systems \cite{Hybertsen1985, Godby1987, Godby1988, Blase1995,vSch2006,Kotani2007,Reining2017,Bruneval2021a,Martin2016}. 

However, there are many cases where the $GW$ approximation is not sufficient. On one hand, the description of QP energies is not always good and certainly worsens in more strongly correlated systems \cite{Romaniello2009,Martin2016}. On the other hand, other quantities, such as satellite features in the electron addition and removal spectra, are often less well described, even in absence of strong correlation \cite{Aryasetiawan1996,Guzzo2011}. Most importantly, $GW$ does not necessarily yield total energies of better quality than currently used density functionals \cite{Bruneval2021b,Stan2009,Caruso2012}. Research on total energies in terms of \GF s goes therefore hand in hand with research on approximations to the self-energy beyond $GW$. The most straightforward way to go would be to explore higher orders in $W$, and important research in this direction is ongoing \cite{Minnhagen1975,Bobbert1994,Shirley1996,Gruneis2014,Hinuma2014,Ren2015,Kutepov2016,Kutepov2017,Pavlyukh2016,Maggio2017,Wang2022}. In many cases it cannot, however, bring a practical solution, since the resulting expressions become quickly very cumbersome and costly, and since perturbation theory will 
diverge when the interaction is too strong.  Therefore, it would be desirable to find an efficient way to terminate the perturbation series. 

In the various possible ways to express the xc energy contribution to the total energy such as using the adiabatic connection fluctuation dissipation theorem \cite{Langreth1975,Langreth1977}, the polarizability plays a key role. This suggests to explore links to other frameworks that are used to access the polarizability, in particular, Time-Dependent DFT (TDDFT) \cite{Gross1984}. Indeed, there is a long, and sometimes very successful, history of attempts to use TDDFT in order to go beyond $GW$ in terms of vertex corrections  based on the xc kernel $f_{\rm xc}$ \cite{DelSole1994,Reining2002,Overhauser1971,Petrillo1988,Mahan1989,Hybertsen1986,Hindgren1997,Schmidt2017,Hung2016,Olsen2019,Chen2015,Tal2021},  the functional derivative with respect to the density of the xc potential $v_{\rm xc}$ of TDDFT \cite{Gross1985}, or related linear response kernels that may be closer to the many-body Green's functions framework \cite{Tokatly2001,Bruneval2005,Gatti2007,Botti2007}. This kind of combination leads to the so-called $G\tilde W$ self-energy, where the Coulomb interaction is screened by a test charge-test electron (TCTE) dielectric function instead of the test charge-test charge (TCTC) one used in the $GW$ approximation \cite{Singhal1976,Hybertsen1987}. Independently of the specific recipe that is used in the various $G\tilde W$ expressions, these approaches replace the complicated exact vertex function $\Gamma$ that depends on three space, spin and time arguments by a two-arguments function $(1-f_{\rm xc}\chi_0)^{-1}$, where $\chi_0$ is an independent-particle polarizability. Therefore, the resulting self-energy is always approximate \cite{Hindgren1997}. Nevertheless, using a $G\tilde W$ self-energy instead of $GW$ often improves the QP energies \cite{DelSole1994,Olsen2019,Tal2021}. At the same time, the idea is much less explored when it comes to total energies \cite{Olsen2019,Hellgren2023}. Moreover, to the best of our knowledge a systematic study for both total energy and spectra that would discern the effect of the replacement of the full $\Gamma$ by a two-arguments vertex on one side, from the effect of approximations to the $f_{\rm xc}$ itself on the other side, is still missing. 

The present work has a focus on the total xc energy, while making a link to other aspects of the \GF\ when interesting. It addresses the following questions: \textit{Could a self-energy with a two-arguments vertex correction, and in particular, a TDDFT-derived one, yield in principle exact results? If yes, how do we have to build the corresponding expressions for the total xc energy? How do widely used approximations impact the results? And what happens to the kinetic energy and to spectra? } As we will show, there are indeed different possibilities to obtain in principle exact expressions for the total xc energy, which are moreover quite robust when widely used approximations are made. Consistent combination of ingredients is a key requirement for this to be true. With these self-energies, the kinetic energy is not exact in principle, but we examine the possibility to use the virial theorem in order to overcome this issue. This allows us moreover to make an interesting comparison to the widely used adiabatic connection approach, which also makes use of the polarizability, but without involving a self-energy. Spectra are also approximate in principle when a two-arguments vertex correction is used, but we find that the $G\tilde W$ results still exhibit improvements over $GW$.

Our investigation and discussion is general, and it is accompanied by an illustration using the exactly solvable symmetric Hubbard dimer at half-filling. The paper is organized as follows: the theoretical background is given in Sec. \ref{theory section}. Theoretical developments made on this basis are presented and discussed in Sec. \ref{theoretical development}. The results for the Hubbard dimer are contained in Sec. \ref{illustrations}. Conclusions are summarized in Sec. \ref{conclusion}.

%% file: theoretical_background_section.tex
\section{Theoretical background \label{theory section} }

\subsection{Total energy and spectral function in terms of the GF}

The ground state total energy $\text{E}_0$ can be expressed in terms of the time ordered GF
\cite{Galitskii1950,Caruso2012}
\begin{multline}
\mathrm{E}_0 = \underbrace{ -i  \lim_{t_2\rightarrow t^+_1}\int dx_1 
   \, \bigg[ -\frac{\nabla^2_{r_1}}{2}+\vext(x_1)\bigg]
   G(x_1,x_1;t_1-t_2)}_{\Ek + \mathrm{E}_{\mathrm{ext}} } \\
  \underbrace{-\frac{i}{2}\lim_{t_2\rightarrow t^+_1} \int{dx_1 \,v_{\mathrm{H}}(x_1)
                  G(x_1,x_1;t_1-t_2)}}_{\EHartree}\\
  \underbrace{- \frac{i}{2}\lim_{t_2\rightarrow t^{++}_1}\int dx_1 dx_3 dt_3\,
      \Sigma_{\mathrm{xc}}(x_1,x_3;t_1-t_3)G(x_3,x_1;t_3-t_2)}_{ \Exc} \, ,
\label{GM equation}
\end{multline}
where $x = (r,\sigma)$ stands for position and spin, and $\displaystyle t^+\equiv \lim_{\eta\to 0^+}\,(t+\eta)$. Here, we have highlighted the different contributions to the total energy, 
namely the kinetic energy $\Ek$, the contribution $\mathrm{E}_{\mathrm{ext}}$ coming from the external 
potential $\vext$, 
the Hartree energy $\EHartree$ given in terms of the Hartree potential $v_{\text{H}}$, and the 
exchange-correlation energy $\Exc$ expressed in terms of the exchange-correlation self-energy $\Sigmaxc$. The last two terms 
compose the interaction energy $\Eint \equiv \EHartree + \Exc$.
Note that here, in the context of MBPT, $\Exc$ refers specifically to the exchange-correlation energy of the Coulomb interaction, 
in contrast to the DFT framework where the xc energy also includes the correlation contribution from the kinetic energy. 
The specific form  Eq. \eqref{GM equation} of the Galitskii-Migdal equation is convenient in order to discuss separately the different contributions to the total energy, and to find specific improvements for each part. While such a strategy does not benefit from error canceling and therefore does not necessarily lead to globally improved results, it helps to obtain deeper insight, and eventually to arrive to the good result for the good reason. \\
The main quantity of interest is $\Sigma_{\text{xc}}$, which can be expressed exactly as
\begin{equation}
\Sigmaxc(1,2) = i \int{d(34) G(1,4) W(3,1^+)  \Gamma(4,2,3)}\, ,
\label{xc self-energy}
\end{equation}
where 
$1 = (x_1,t_1)=(r_1,\sigma_1,t_1)$ stands for position, spin and time. The screened Coulomb interaction 
$W$ is given by
\begin{equation}
    W(1,2) = v_c(1,2) + \int d(34)\,v_c(1,3)\chi(3,4)v_c(4,2) \, , 
    \label{eq:W}
\end{equation}
with $v_c$ the bare Coulomb interaction $v_c(1,2) = \delta(t_1-t_2) \frac{1}{|r_1-r_2|}$ and $\chi$ the reducible polarizability. The main complication stems from the
vertex function $\Gamma$, defined as
\begin{equation}
\Gamma(4,2,3)  = \delta(4,3)\delta(2,3) + \frac{\delta \Sigma_{ \text{xc} }(4,2)}{\delta v_{ \text{cl} }(3^{++},3^+) }\,,
\label{eq:vertex}
\end{equation}
where $v_{\rm cl} = v_{\text{H}} + \vext $ is the total classical potential. Since $\Gamma$ is in turn determined by the self-energy, in most cases it cannot be expressed in a closed form nor calculated exactly. To lowest order in the Coulomb interaction $\Gamma(4,2,3) \approx \delta(4,3)\delta(2,3)$. 
Corrections to this result are called vertex corrections. 
 Neglecting vertex corrections one obtains the $GW$ approximation, where $\Sigmaxc(1,2) = i G(1,2)W(2,1^+)$\cite{Hedin1965}.

Once the self-energy is determined in the chosen approximation, the \GF\ is obtained by solving the 
Dyson equation
\begin{equation}
 G(1,2) = G_0(1,2) + \int{d(34)} \,G_0(1,3)\bigg( \vH(3,4) +\Sigmaxc(3,4)\bigg)G(4,2)\,, 
\end{equation}
where $G_0$ is the non-interacting \GF . 
Finally, the resulting \GF\ can be used to calculate $\text{E}_0$ from Eq. \eqref{GM equation}
or to evaluate the spectral function from the frequency Fourier transform of $G$,

\begin{equation}
 A(x,x,\omega) = \frac{1}{\pi} |\Im\big(G(x,x;\omega) \big)|\,.
 \label{spf in terms of the 1-GF}
\end{equation}

\subsection{Interaction energy in terms of the polarizability}
%\subsection{Coulomb interaction energy in terms of the polarizability}
Our focus is to find accurate expressions for the interaction energy
$\Eint$. For this purpose,
it is useful to express it
in terms of the reducible polarizability $\chi$ \cite{Hellgren2023}.

For a system with $N$ electrons in its ground state, the interaction energy is given 
by the expectation value of the Coulomb interaction operator $\hat{V}$ in the many-body ground state  $\ket{N_0}$,
\begin{equation}
\Eint = \braket{N_0|\hat{V}|N_0}= \frac{1}{2} \int{dx_1dx_2 }\,\vcoul(x_1,x_2)  \bra{N_0} \hat{\psi}^\dagger(x_2) \hat{\psi}^\dagger(x_1)  \hat{\psi}(x_1) \hat{\psi}(x_2) \ket{N_0}\,,
\label{interaction energy}
\end{equation}
where $\hat{\psi}$ and $\hat{\psi}^\dagger$ are the annihilation and creation field operators, respectively.
On the other hand, the reducible polarizability $\chi$ is defined as
\begin{multline}
\chi(x_1,t_1;x_2,t_2) = - i G(x_1,t_1;x_1,t_1^+)G(x_2,t_2;x_2,t_2^+) \\ - i \bra{N_0} \hat{T}\big[ \hat{\psi}^\dagger(x_1,t^+_1)  \hat{\psi}(x_1,t_1)  \hat{\psi}^\dagger(x_2,t_2^+)\hat{\psi}(x_2,t_2)\big] \ket{N_0} \,. 
\label{chi in terms of the 2-GF}
\end{multline}
In the limit $t_2 = t_1^{++}$,
%\begin{align}
%G_2&\begin{multlined}[t](x_1,t_1,x_2,t_1^+;x_1,t_1^+,x_2,t_1^{++}) = \\- \bra{N} \hat{\psi}^\dagger(x_2) \hat{\psi}(x_2)  \hat{\psi}^\dagger(x_1) \hat{\psi}(x_1) \ket{N} \end{multlined} \\
%%&=\begin{multlined}[t] - \bra{N}\hat{\psi}^{\dagger}(x_2) \hat{\psi}(x_1) \ket{N} \delta(x_2-x_1) \\
%% - \bra{N} \hat{\psi}^\dagger(x_2) \hat{\psi}^{\dagger}(x_1) \hat{\psi}(x_1)\hat{\psi}(x_2) \ket{N} \end{multlined}\\ 
%&=\begin{multlined}[t] -n(x_1,x_2) \delta(x_2-x_1) \\
 %- \bra{N} \hat{\psi}^\dagger(x_2) \hat{\psi}^{\dagger}(x_1) \hat{\psi}(x_1)\hat{\psi}(x_2) \ket{N} \end{multlined} 
%\end{align}
%\begin{align}
%G_2(x_1,t_1,x_2,t_1^+;&x_1,t_1^+,x_2,t_1^{++}) = - \bra{N} \hat{\psi}^\dagger(x_2) \hat{\psi}(x_2)  \hat{\psi}^\dagger(x_1) \hat{\psi}(x_1) \ket{N} \\
%&= -\rho(x_1,x_2) \delta(x_2-x_1) 
% - \bra{N} \hat{\psi}^\dagger(x_2) \hat{\psi}^{\dagger}(x_1) \hat{\psi}(x_1)\hat{\psi}(x_2) \ket{N} 
%\end{align}
\begin{multline}
 \chi(x_1,t_1;x_2,t_1^{++}) = -i G(x_1,t_1;x_1,t_1^+)G(x_2,t_1^{++};x_2,t_1^{+++})\\
 -i \braket{N_0|\hat{\psi}^{\dagger}(x_2) \hat{\psi}(x_1) |N_0}\delta(x_2-x_1) 
 -i \braket{N_0| \hat{\psi}^\dagger(x_2) \hat{\psi}^{\dagger}(x_1) \hat{\psi}(x_1)\hat{\psi}(x_2) |N_0}  \\
 = i\, n(x_1)n(x_2) - i\, \rho(x_1,x_2) \delta(x_2-x_1) 
-i  \braket{N_0| \hat{\psi}^\dagger(x_2) \hat{\psi}^{\dagger}(x_1) \hat{\psi}(x_1)\hat{\psi}(x_2) |N_0} \,,
 \label{chi in terms of the field operator}
\end{multline}
where we used the anticommutation relation $\{\psi(x_2),\psi^\dagger(x_1) \} = \delta(x_2-x_1)$, and where we introduced the one-body reduced density-matrix $\rho(x_1,x_2) =  \bra{N_0}\hat{\psi}^{\dagger}(x_2) \hat{\psi}(x_1) \ket{N_0} = -i G(x_1,t,x_2,t^+)$, with the electron density $n(x_1)=\rho(x_1,x_1) $. The last term in Eq. \eqref{chi in terms of the field operator} enters the definition of the 
interaction energy in Eq. \eqref{interaction energy}. The interaction energy can therefore be expressed in terms of 
the polarizability $\chi$ as
\begin{multline}
\Eint=  \frac{1}{2} \int{dx_1dx_2} \vcoul(x_1,x_2) n(x_1) n(x_2) \\
 + \frac{i}{2} \int{dx_1dx_2 \vcoul(x_1,x_2) \chi(x_1,t_1;x_2,t_1^{++})}  
- \frac{1}{2} \int{dx_1dx_2 \vcoul(x_1,x_2) \rho(x_1,x_2) \delta(x_2-x_1)}\,,
%=\EHartree + \Exc,
\label{Coulomb interaction as a function of chi and delta function}
\end{multline}
where the first term is the Hartree energy and the last two terms are the exchange-correlation energy. This formulation of $\text{E}_{\rm xc}$ is not directly suitable for practical purposes, since it consists of terms containing a divergence that cancels in the sum. It is, however, a good starting point for the developments in the next section. 
%The last term in equation \ref{Coulomb interaction as a function of chi and delta function} corresponds to the Hartree energy, as the 1-GF at equal time, space, and spin $G(x,t;x,t^+)$ is equivalent to the density $n(x)$. The remaining two parts represent the xc energy ($\text{E}_{\text{xc}}$).  
%It is important to note that the second term in the equation diverges due to the presence of $\delta(x_2-x_1)$, 
%indicating the existence of a cancellation originating from another divergent term contained in $\chi$.

%% file: theoretical_development_section.tex
\section{Theoretical developments \label{theoretical development} }

\subsection{A freedom of choice \label{diversity of choices for chi0} }
\label{subsec:freedom}
%The expression of the exchange-correlation energy in terms of the polarizability given by the second line of 
%Eq.\eqref{Coulomb interaction as a function of chi and delta function} represents our starting point. 
%It is important to notice that this expression diverges due to the presence of $\delta(x_2-x_1)$, 
%indicating the existence of a cancellation 
%originating from another divergent term contained in $\chi$. 
In order to eliminate the problematic last term
%divergent term $\braket{N|\hat{\psi}^{\dagger}(x_2) \hat{\psi}(x_1) |N} \delta(x_{2}-x_{1})$ 
in Eq. \eqref{Coulomb interaction as a function of chi and delta function}, we introduce a
generalized independent-particle polarizability defined as $\chi_0(1,2)\equiv -i\bar G(1,2^+)\bar G(\bar 2,1^+)$. 
Its time diagonal is
%We also show that the choice of $\chi_0$ is flexible. 
\begin{align}
\chi_0(x_1,t,x_2,t^{++})  &= -i\bar G(x_1,t;x_2,t^{+++})  \bar G(x_2,t^{++};x_1,t^+)  \\
&= \bar \rho(x_1,x_2) \bigg( -i\bra{\bar{N}_0 } \hat{\psi}(x_2)  \hat{\psi}^\dagger(x_1) \ket{\bar{N}_0 } \bigg) \\
&= -i \bar \rho(x_1,x_2)  \bra{\bar{N}_0 }\delta(x_2-x_1)  - \hat{\psi}^\dagger(x_1)\hat{\psi}(x_2)  \ket{\bar {N}_0 } \\
&= -i\bar \rho(x_1,x_2)\delta(x_2-x_1)+ i\bar \rho(x_1,x_2)\bar \rho(x_2,x_1) \, ,
\label{generalized chi0}
\end{align}
where $\ket{\bar{N}_0}$ is the many-body ground state corresponding to a system that could be the true interacting system or an auxiliary interacting or non-interacting system. $\bar G$ and $\bar \rho$ are the corresponding \GF\ and the corresponding density matrix, respectively.
%This generalization will show 
%the freedom of choice in the independent particle
%starting point, that will be crucial both to produce a general recipe for an exact total 
%energy expression and to guide the design of efficient approximations. 

In the last term of Eq. \eqref{Coulomb interaction as a function of chi and delta function}, only the diagonal of the density matrix is needed. In order to replace this term, we can therefore consider all systems that yield the exact density $\bar \rho(x,x)=n(x)$, such as the true interacting system, or the Kohn-Sham auxiliary system. This leaves considerable freedom, which we can use to derive different exact expressions for $\Eint$ and to design efficient approximations. Indeed, when $\bar \rho(x,x)=n(x)$ we have, from Eq. \eqref{generalized chi0}
\begin{equation}
 n(x_1)\delta(x_2-x_1) = i\chi_0(x_1,t;x_2,t^{++}) +\bar \rho(x_1,x_2)\bar \rho(x_2,x_1)\,,
 \label{mb density delta}
\end{equation}
which, replaced in Eq. \eqref{Coulomb interaction as a function of chi and delta function}, yields
\begin{align}
 \text{E}_{\text{xc}} &= \begin{multlined}[t] 
  - \frac{1}{2} \int{dx_1dx_2 \vcoul(x_1,x_2) \bar \rho(x_1,x_2)\bar \rho(x_2,x_1)}
  \\ 
  + \frac{i}{2} \int{dx_1dx_2} \vcoul(x_1,x_2) \bigg( \chi(x_1,t_1;x_2,t_1^{++})
   - \chi_0(x_1,t_1;x_2,t_1^{++}) \bigg) \end{multlined} \label{Exc generalized integration} \\
 &= \Exb + \Ecb = \Exb + \Ec + \left(\Ex-\Exb\right) \,.
 \label{xc general formula}
\end{align}
The first term in Eq. \eqref{Exc generalized integration} is $\Exb$, 
the exchange energy corresponding to $\ket{\bar{N}_0}$.
Since the derivation shows that the sum of all terms is the exact exchange-correlation energy, the second term in Eq. \eqref{Exc generalized integration} $\Ecb$ contains the exact correlation energy plus a correction that compensates the error of $\Exb$ with respect to the exact exchange energy $\Ex$.  
%\emph{ The crucial observation here is that the term $n(x_1,x_2)\delta(x_2-x_1) = n(x_1,x_1)\delta(x_1-x_2)$ contains the exact density. Therefore, we can alternatively obtain the exact $\text{E}_{\text{xc}}$ by utilizing the $KS$ non-interacting polarizability ($\chi_0^{KS}$), since the KS scheme is designed to reproduce the exact density. In fact, any other choice of $\chi_0$ can also be employed as long as it accurately reproduces the exact density}.
\emph{It is crucial to note that one can use any system defined by a ground state $\ket{ \bar{N}_0 }$, as long as this system yields the exact density: this will yield the exact Coulomb interaction energy, although $\bar \rho$ is not the density matrix of the true interacting system.} Two most obvious choices are either the true many-body (MB) system with $\bar G=G$, which leads to $\chiomb \to -iGG$ and $\bar \rho(x_1,x_2)=\rho(x_1,x_2)$ the true density matrix, or the Kohn-Sham (KS) system with $\chi_0\to\chioks\equiv -i\Gks\Gks$ the independent-particle polarizability built with the Kohn-Sham Green's function, and $\bar \rho\to \rho^{\text{KS}}$ the KS density matrix.

\subsection{Exact exchange-correlation energy from approximate self-energies \label{xc theoretical development}}

%We have expressed, in the previous section, the exchange-correlation energy in terms of polarizability, that permits us to consider 
%different starting points. Our aim is now to exploit this freedom in a formula that permits us to evaluate the self-energy and, via
%the GM and the Dyson equation, both the total energy and the spectral function. 
Our next goal is to make a self-energy appear in the expression of $\Exc$. 
To this aim, we rewrite Eq. \eqref{Exc generalized integration} as
\begin{align}
 \Exc&= \frac{1}{2}\int dx_1d2\, \bar G(1,2^+)\bar G(2,1^{+} )\vcoul(2,1) + 
      \frac{i}{2} \int dx_1 d3\, \vcoul(3,1)\left[\chi(1,3^{++})-\chi_0(1,3^{++}) \right]\label{eq:Ex+Ec} \\
 &= \begin{multlined}[t]\frac{1}{2}\int dx_1 d2 \, \bar G(1,2^+)\bar G(2,1^{+ })\vcoul(2,1) \\   
        + \frac{i}{2} \int dx_1(234)\, \chi_0(1,2)\left[\vcoul(2,4)+\fxcb(2,4) \right]\chi(4,3^{++}) \vcoul(3,1)\,,\end{multlined} \label{chitddft}
\end{align}
where we have introduced the generalized exchange-correlation 
kernel $\fxcb$ that, once a choice for $\chi_0$ is made, is defined from the Dyson-like equation
\begin{equation}
\chi(1,2) = \chi_0(1,2) + \int d(34) \chi_0(1,3)\bigg( \vcoul(3,4) + \fxcb(3,4) \bigg) \chi(4,2)\,,  
\label{eq:deffxc}
\end{equation}
 keeping in mind that $\chi$ is always the exact reducible polarizability. When $\chi_0$ is chosen to be the KS independent particle polarizability, $\bar f_{\rm xc}= f_{\rm xc}$, the xc kernel of 
 linear response TDDFT\cite{Gross1985},  but, as pointed out above, other choices are possible.\footnote{Note that here we have given the equations in terms of time-ordered quantities, whereas TDDFT is usually causal. One has to pay attention to be consistent when combining the GFFT and TDDFT frameworks in practice.} By using the definition of $\chi_0$, given in the beginning of \ref{subsec:freedom}, Eq. \eqref{chitddft} can be written as 
\begin{align}
 \Exc&=
%\begin{multlined}[t]\frac{1}{2}\int dx_1 d2 \, \bar G(1,2^+)\bar G(2,1^{+})\vcoul(2,1) \\         + \frac{1}{2} \int dx_1(234)\, \bar G(1,2^+) \bar G(2,1^{+}) \left[\vcoul(2,4)+\fxcb(2,4) \right]\chi(4,3^{++}) \vcoul(3,1)\end{multlined} \label{pluses in chi}\\
 \int{dx_1 d2\, \bar G(1,2^+) \bar G(2,1^{+}) \bigg( \vcoul(2,1) + \int{d(34)\, \big( \vcoul(2,4) + \fxcb(2,4) \big) \chi(4,3^{++}) \vcoul(3,1) } } \bigg)     \\  
 &= \int{dx_1 d2\, \bar G(1,2) \bar G(2,1^{++}) \bigg( \vcoul(2,1^+) + \int{d(34)\, \big( \vcoul(2,4) + \fxcb(2,4) \big) \chi(4,3^{++}) \vcoul(3,1^+) } } \bigg)     \\  
 &= \frac{1}{2}\int dx_1d2 \, \bar G(1,2) \Wbt(2,1^+)  \bar G(2,1^{++})\,,
 \label{Wtcte}
\end{align}
where we have defined the  generalized TCTE screened Coulomb interaction\footnote{The double infinitesimals in $\chi(4,3^{++})$ do not change the spectrum of $\Wbt$, but we keep them here explicitly since they give a straightforward prescription for the contour integral in frequency space yielding $E_{\rm xc}$.}
\begin{equation}
\Wbt(2,1) = \vcoul(2,1) +  \int{d(34)} \bigg( \vcoul(2,4) +\fxcb(1,4)\bigg)\chi(4,3^{++}) \vcoul(3,1)\,. 
\end{equation} 
In this way, the \textit{exact} exchange-correlation energy takes a form analogous to the last term in the Galitskii-Migdal expression Eq. \eqref{GM equation}:
\begin{equation}
 \Exc= -\frac{i}{2}\int dx_1 d2 \, \Sigmaxcb(1,2)\bar G(2,1^{++}) \,,
 \label{xc GM expression}
\end{equation}
with an exchange-correlation self-energy
\begin{equation}
    \Sigmaxcb(1,2)\equiv i\bar G(1,2)\bar {\tilde W}(2,1^+)\,.
    \label{eq:tcte-sigma}
\end{equation}
%\textcolor{blue}{It is worth noting that the $\Sigmaxc$ resulting from the derivations starting from Eq.\eqref{eq:Ex+Ec} incorporates infinitesimal pluses in $\chi$, as seen in Eq.\eqref{pluses in chi}. These pluses are of utmost importance for obtaining the correct solution. Neglecting them can alter the contour direction and consequently yield an incorrect result. Therefore, it is imperative to account for these infinitesimal pluses in the calculations.}  

The important point to stress here is the fact that the \textit{exact} $\Exc$ is obtained with an \textit{approximate}
self-energy Eq. \eqref{eq:tcte-sigma}.
This approximation is often called $G\tilde W$. 
% with respect to the exact one, the vertex correction beyond 
%the GW expression is approximated.}
It is usually derived \cite{DelSole1994} by replacing  $\Sigmaxc(4,2)$ in the functional derivative in Eq. \eqref{eq:vertex} with a local $\delta(4,2)\bar v_{\rm xc}(4)$. Most often, $\bar v_{\rm xc}\equiv v_{\rm xc}$, the KS xc potential of TDDFT, is chosen and the resulting $f_{\rm xc}$ is approximated, for example, in the adiabatic local density approximation. In our derivation, $\fxcb$ does not have to be  a functional derivative, since it is defined by Eq. \eqref{eq:deffxc}, which 
generalizes the definition of $\Wbt$. 
This gives a rigorous foundation to attempts to use $\fxcb$ other than approximate TDDFT ones in order to approximate vertex corrections to the self-energy, in particular, the so-called nanoquanta kernel and approximations to it:\cite{Tokatly2001,Bruneval2005,Gatti2007,Botti2007,Reining2002,Sottile2003,Sharma2011,Adragna2003,Marini2003,Rigamonti2015} the only requirement is that $\chi_0$ corresponds to the correct density. 
It should, however, be noted that a $\fxcb$ fulfilling 
Eq. \eqref{eq:deffxc} does not necessarily exist for every $\chi_0$. We will give an illustration below in the Hubbard dimer.

The important message of this section is that \emph{the exact exchange-correlation energy can be obtained with an approximate 
self-energy $\bar \Sigma_{\rm xc}$ and with an approximate Green's function $\bar G$ which is not the solution of the Dyson equation using 
$\Sigmaxcb$, but which has been chosen from the beginning.  
The two important requirements are consistency of the ingredients used in Eq. \eqref{xc GM expression}, and the fact that they stem from 
the real or from an auxiliary system yielding the exact density.} In the following, we will call this a \textit{\consistent{} scheme}, 
as opposed to a non-\consistent{} scheme where different \GF s are used in $\Exc, \Sigmaxcb, \Wbt$. 
%In \cite{BrunevalPRL} it was anticipated that non-locality corrections to the vertex may integrate to zero in the correlation energy \textcolor{red}{Actually, I cannot find this statement anymore. Can anyone of you find it in Fabien's article? }. 
Here, we have shown that there is more than one possible consistent choice, which may help to design efficient approximations.

\subsection{The kinetic energy 
%Linearized Dyson Equation (LDE) and virial theorem 
\label{kinetic energy theoretical development} }
The TCTE screened self-energy $\bar \Sigma_{\rm xc}$ does in general not correspond to the exact self-energy, and therefore one does not have access to the exact \GF\ nor to the exact density matrix.
%, nor to the \GF\ resulting from the Dyson equation using $\bar \Sigma_{\rm xc}$ exact. 
As a consequence, the kinetic energy $\Ek$ cannot be computed exactly. However, with the exact Coulomb interaction energy $\Eint$ at hand, this problem can in principle be overcome by using the virial theorem for the electron system
\cite{Levy1985,Jiang2020},
\begin{equation}
 2\Ek + \Eint = \int{d^3{\bf r} \, n({\bf r}) {\bf r}\cdot\nabla \vext(r) } 
 \equiv S_{VT}\,.
 \label{VT}
\end{equation}
Using the virial theorem requires in principle to know the exact density. 
This is not an additional requirement here, since it was already assumed throughout the above derivations. Moreover, research in the framework of DFT shows that errors induced by approximate functionals are often predominantly due to the form of the functional, whereas in many cases errors due to an approximate density are small \cite{Kim2013}. Therefore, the use of the virial theorem is a promising route to take when, as it is the case here, one can expect to access the interaction energy with good accuracy.

\subsection{Comparison to the adiabatic connection \label{AC}}
Finally, it is interesting to compare our equations to
the adiabatic connection (AC) approach \cite{Langreth1975,Langreth1977,Ren2012}. In principle, this approach yields the exact full correlation energy, which encompasses correlations arising from both kinetic and Coulomb interaction energies, as well as the difference between the exchange energy calculated with the true and the KS density matrix, respectively.

Also in this approach, the correlation energy is expressed in terms of $\chi$ and $\chi_0$, but with an integration over a coupling constant $\lambda$ that scales the Coulomb interaction and modifies $\vext$ such that the density remains constant, 
%\cite{Jiang2020},
\begin{equation}
 \text{E}^{\text{full}}_{\text{c} } =\frac{i}{2}  \int_0^1{ d\lambda}\int{dx_1dx_2}\, \vcoul(x_1,x_2) \bigg( \chi^\lambda(x_1,t_1;x_2,t_1^{++}) - \chioks(x_1,t_1;x_2,t_1^{++})  \bigg)\,.
 \label{adiabatic connection}
\end{equation}
Since the structure of the expression is the same as that of $\Ecb$ in Eq. \eqref{xc general formula}, one can express also the AC result in terms of an effective self-energy,

\begin{equation}
\text{E}^{\text{full}}_{\rm c} = -\frac{i}{2} \int{dx_1d3 \,\Sigma^{\text{eff}}_{\text{c} }(1,3) \Gks(3,1^{++})},
\end{equation}
where
\begin{equation}
  \Sigma^{\text{eff}}_{\rm c}(1,3) = i\Gks(1,3) \int_0^1{ d\lambda\, \tilde{W}_{\rm pol}^{\lambda}(3,1^+) }\,,
\end{equation}
with $\tilde{W}_{\rm pol}^{\lambda}$ the polarization contribution to the $\lambda$-dependent KS TCTE screened interaction. 
$ \Sigma^{\text{eff}}_c$ is an effective correlation self-energy that contains kinetic and interaction contributions. Here, we have worked with the KS scheme, since the AC expression is often (though not always, see, e.g., \cite{Maggio2017}) used in the framework of KS-DFT. Analogous expressions are obtained for other allowed choices of $\chi_0$, e.g., stemming from a generalized KS scheme. Of course, this self-energy yields the exact full correlation energy, while it is not meant to be used in a Dyson equation to yield the \GF . \\

It is interesting to compare the errors to be expected in practice from the AC approach on one side, and, with the errors of the approach discussed here, i.e., the combination of the calculation of $\Eint$ plus the use of the virial theorem. For this estimate, we suppose the virial term $S_{\rm VT}$ in Eq. \eqref{VT} to be known with an error that is negligible with respect to the error $\Delta \Eint$ stemming from approximations to $\chi$.  This is consistent with the fact that we suppose the density to be known with good accuracy. Using the virial theorem $2 \Ek + \Eint =S_{\rm VT}$, the error in the kinetic energy will be  $\Delta \Ek = -\frac{\Delta \Eint }{2}$, leading to a total error of $\Delta \text{E} =+\frac{\Delta \Eint}{2}$. 
In the case of the AC, the error is determined entirely by the integral over response functions Eq. \eqref{adiabatic connection}. Since the non-interacting $\chi_0$ is subtracted, it is reasonable to suppose that the dominant contribution is linear in $\lambda$. Evidence that this is true can be found for small systems in Ref. \cite{Savin2001}. Assuming linearity in $\lambda$, one obtains the \emph{same} error $\Delta \text{E}=+\frac{\Delta \Eint }{2}$ as in our alternative scheme. Whether higher orders in $\lambda$ will rather reduce or increase this result depends on whether $\chi_{\lambda}$ is convex or concave. In any case, this discussion suggests that similar errors are to be expected, while the $\lambda$-integration is avoided in the approach using the virial theorem.

\subsection{Shortcomings of the TCTE self-energy}
\label{subsec:shortcomings}
While different flavors of the TCTE screened $G\tilde W$ self-energies  yield the exact xc energy, 
they will in general not yield the correct spectral function calculated from the solution of the Dyson equation. 
One may expect some improvement with respect to the $GW$ approximation for the quasiparticle (QP) energies, 
since the use of $\bar f_{\rm xc}$, which is negative, reduces the  polarization contribution and therefore 
approximates one important effect of the full vertex corrections, which is to reduce self-polarization \cite{Romaniello2009}. 
However, one may expect that it will not be sufficient to bring significant correction to the satellites, 
which are in general poorly described by the $GW$ approximation. One reason lies in the fact that the $GW$ self-energy is of first order in $W$, but used in the solution of the Dyson equation plus infinite order. Another reason is the following: the poles of the exact 
Green's function are the total energy differences $\pm (E_{N\pm 1,s}-E_N)$, where $N$ is the particle number and $s$ 
labels a ground ($s=0$) or excited state $s$. This can be written as $\pm (E_{N\pm 1,s}-E_{N\pm 1,0}) \pm (E_{N\pm 1,0}-E_N)$, 
i.e., the excitation energy of the $N\pm 1$-electron system plus the chemical potential for electrons or holes.
This means that satellites of the QP, that lies at the respective chemical potential, must be found at a distance equal to the
excitation energies of the $N\pm 1$-electron system, and not, as it would be the case in the $GW$ approximation for small 
systems with a discrete spectrum,  at a distance close to the excitation energies of the $N$-electron system (plus differences in input and output QP energies when $G$ is not calculated self-consistently). 
%\textcolor{red}{Attention, en GW c'est la partie imaginaire de sigma qui se trouve a cette distance, tandis que la
%solution de l'equation de dyson induit un shift. Comme nous voyons pour le hubbard dimer fig. 9, ce shift est zero 
%quand on a peu de pics.}
This shortcoming cannot be overcome by a $\fxc(\omega)$ that depends on a single frequency 
and multiplies $\chi(\omega)$ in frequency space: such a structure cannot shift the poles of $\chi(\omega)$.
This could only be achieved by a frequency integration, as it is the case when the true three-times vertex correction is used. 
One should therefore at best expect corrections of the intensities of the satellites when moving from $GW$ to $G\tilde W$.

%% file: illustrations_new.tex
\section{Illustrations \label{illustrations} }

In order to illustrate our main findings and suggestions, we will use a simple exactly solvable model, the symmetric Hubbard dimer\cite{Romaniello2009,Romaniello2012,Carrascal2015,Aryasetiawan2002,Coveney2023}. Its hamiltonian reads\cite{Hubbard1963,Hubbard1964}
%Exactly solvable models are crucial for theoretical investigations and illustrations. Having the exact result at hand, we are able to explore any approximation. In the present article, we use for illustrations the widely used Hubbard dimer model, which is made up of two equivalent sites, each has one single orbital and on-site Coulomb interaction U. The orbitals are localized, since the Hubbard model is performed initially to deal with strongly correlated physics, therefore the electron can go to the nearest neighboor site with a hopping integral $\text{t}$. The model hamiltonian in second quantization is written as follows,
\begin{equation}
\hat{H} =  \sum_{i,\sigma}\epsilon_0 \hat{n}_{i\sigma} -\sum_{<i,j>,i\ne j,\sigma} t\hat{c} ^{\dagger}_{i\sigma} \hat{c}_{j\sigma} + U\sum_{i} \hat{n}_{i\uparrow} \hat{n}_{i\downarrow}.
\label{Hubbard hamiltonian}
\end{equation}
where $i,j$ denote the sites $1,2$, the spin $\sigma=\uparrow,\downarrow$, the external on-site potential is
$\epsilon_0$, and 
$t$ is the hopping that is linked to the kinetic energy. 
$U$ is the onsite Coulomb repulsion, and
$\hat{n}_{i\sigma}=\hat{c}^{\dagger}_{i\sigma}\hat{c}_{i\sigma}$ is the particle number operator, where $\hat c$ and $\hat c^{\dagger}$ annihilate and create a fermion, respectively. Using this simple model allows us to explore the full range of correlation, which can be quantified by the ratio $U/t$, and to have an unambiguous benchmark. We will use it at half filling, i.e., with two electrons, which yields non-trivial electron removal and addition features, and we will set $U=$4 eV throughout the illustrations. One limitation of the model is the fact that the density is trivial and always exact in all methods that conserve symmetry and particle number. Since in the present work we suppose to know the exact KS ingredients, this is not a main drawback. Moreover, asymmetry in the potential removes degeneracy and therefore has a tendency to decrease correlation effects. The symmetric dimer is therefore the most critical test case. Exploring density-driven errors would be interesting, but beyond the scope of this work.

\begin{figure*}[ht]
\centering
\includegraphics[width=1\textwidth]{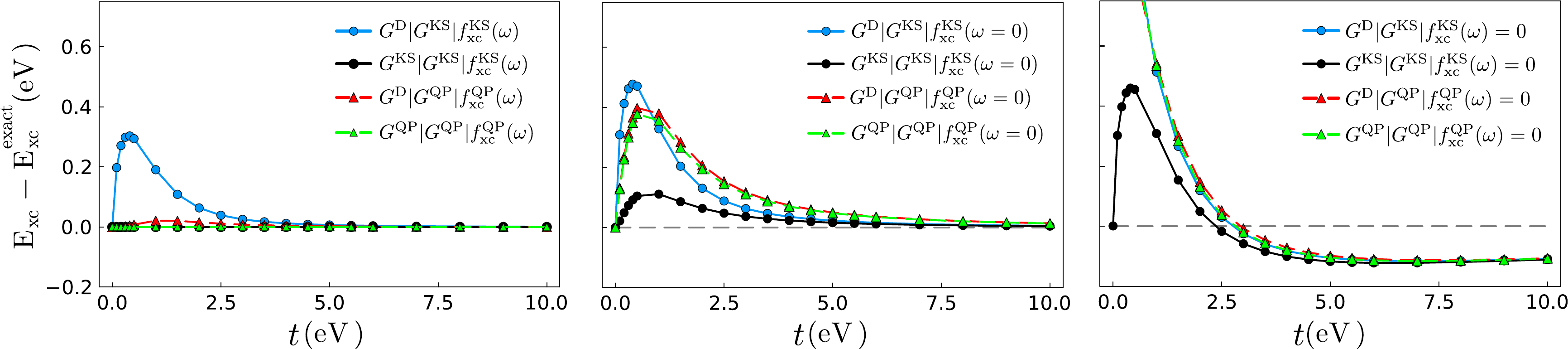} 
\caption{Symmetric Hubbard dimer at half filling and $U=4$ eV: error of the exchange-correlation energy as a function of the hopping $t$. $\Exc$ is obtained from $\Exc=-\frac{i}{2}\int \Gout\Sigmaxc[\Gin]$ for different $\Gin,\Gout$, which are, respectively, the input \GF\ used to build the self-energy, and the \GF\ that is usually the output of the Dyson equation, but for which we have more options here.  $\Sigmaxc$ is a $G\tilde W$ self-energy, built with $\Gin$ and using the consistently chosen xc kernel.
For a compact notation, we denote this by $\Gout|\Gin|\fxcb(\omega)$. 
The black and sky blue solid lines with dot markers result from a self-energy built with KS ingredients and integrated, respectively, consistently with $\Gout=\Gin=\Gks$ or, inconsistently, with the $\Gout=\Gd$ resulting from the Dyson equation. 
The red and green dashed lines with triangle markers result from a self-energy built with QP ingredients and integrated, respectively, consistently with the $\Gout=\Gin=\Gqp$ or, inconsistently, with the $\Gout=\Gd$ resulting from the Dyson equation. Left panel: results using the exact consistent $\fxcb(\omega)$.
Middle panel: results using the adiabatic approximation $\omega=0$ for $\fxcb$. Right panel: results obtained by neglecting $\fxcb$ completely, which corresponds to a $GW_0$ approximation, where $W_0$ is calculated in the RPA and $G=\Gin$.
\label{consistent xc} }
\end{figure*}

The exact analytical expressions for the time-ordered Green's function and self-energy are given in \appendixinfo. For the approximate Green's functions we have solved the Dyson equation numerically. Our code uses retarded quantities \cite{Spataru2004,Honet2022} which is more reliable, since the numerical results of the time-ordered calculations suffer from instabilities for some approximations in the small range of $t$ ($t \rightarrow 0$). Computational details are given in \appendixinfo.
%However, when it comes to the 1-GFs obtained from the Dyson equation and the total energies, we have relied on numerical calculations. The analytical results provided in the appendices are given in terms of time-ordered quantities.
%In the numerical calculations, we have used retarded quantities instead of time-ordered ones. The only difference is that we have replaced the $-i\eta$ term in the poles with $+i\eta$ . This modification is made to prevent any overlap between negative and positive peaks, particularly in cases of strong correlation when $t \rightarrow 0$.} 

%In this section, we split the total energy and spectra results in two different subsections \ref{Etot section} and \ref{spectra section}.

As worked out in Subsec. \ref{subsec:freedom}, different choices for 
$\chi_0$ are possible. The simplest choice is to build $\chi_0$ with KS Green's functions. In this case, the corresponding xc kernel $\fxcks$ is the one defined in TDDFT. In the symmetric Hubbard dimer the KS xc potential is a number that we set by constraining the highest occupied level (HOMO) energy of the KS system to yield the exact ionization potential. In this way we obtain the KS Green's function and $\chioks$, and subsequently $\fxcks$ by inversion of Eq. \eqref{eq:deffxc}. 
This inversion is not unique in the symmetric Hubbard dimer, because both the exact $\chi$ and $\chioks$ have only one non-zero element, which is the antibonding/antiboding one (see \appendixinfo). Therefore, as already pointed out in \cite{Aryasetiawan2002}, only the antibonding/antibonding matrix element of the resulting $\fxc$ is defined. The other elements are arbitrary, but their choice has no impact on the results, since $\fxc$ appears only in the combination $\chi_0\fxc\chi$.  

Another natural choice would be to use the exact Green's function $G$ to build $\chi_0$, since it also yields the exact density, as required. However, interestingly there is no solution to the inversion of Eq. \eqref{eq:deffxc} in this case. The reason is that also the bonding/bonding element of this $\chi_0 = -i GG$ is non-vanishing. Further analysis shows that this stems from the satellite contributions to $G$, which are not canceled by proper vertex corrections. This is a nice illustration for one of the problems of this ill-behaved polarizability which also, for example, does not fulfill the $f$-sum rule \cite{Holm1998}. We will instead use $\chioqp \equiv-i\Gqp \Gqp$. It is built with the QP approximation $\Gqp$ to the exact $G$, where satellites are neglected and the remaining intensities normalized to 1. This can be seen as a realization of a generalized KS Green's function, stemming from a potential that is non-local in space but instantaneous in time. Such a potential can lead to accurate QP energies \cite{Perdew2017}, but not to satellites. In a real material, the widely used hybrid functionals \cite{Heyd2003} fall into this class. Also many scalar long-range kernels are designed to be used on top of a $\chioqp$. It should again be stressed that both kernels, whether the one of the KS or the one of the QP scheme, can be called ``exact'', as long as they are used consistently in conjunction with $\chi_0$ built with the corresponding Green's functions.

\subsection{ Results using exact xc kernels}

In the following we will focus on the results obtained with the two kernels $\fxcks(\omega)$ and $\fxcqp(\omega)$, without approximating them further. This will allow us to 
illustrate the effect of using an $\fxc$ to simulate the full three-argument vertex of many-body perturbation theory, without further approximations.

\subsubsection{Exchange-correlation energy}

First, let us examine the xc contribution to the total energy, given by Eq. \eqref{xc GM expression} $\Exc=-\frac{i}{2}\int \bar G\Sigmaxcb$. As pointed out above, here $\bar G$ should \textit{not} be the exact Green's function nor the one resulting from the Dyson equation with $\Sigmaxcb$, which we will call $\Gd$ in the following, but $\bar G$, which is the one used to build the $G\tilde W$ self-energy $\Sigmaxcb$. This point is important since in practical applications, using $\Gd$ would often seem to be a natural choice, being the best available Green's function, i.e. the one closest to the exact $G$. We will therefore compare these choices in the following, by evaluating $\Exc=\int G^{ \text{out} }\Sigmaxcb[G^{\text{in}}]$. Here, $\Gin$ is the input \GF\ used to build the $G \Wbt$ self-energy $\Sigmaxcb$, and the $\Gout$  is either the output of the corresponding Dyson equation $\Gd$, or equal to $\Gin$. In all cases, $\Sigmaxcb$ is  built with the xc kernel $\fxcb$ that is consistent with $\Gin$.

For a compact notation, we use $\Gout|\Gin|\fxcb(\omega)$. For example, $\Gd|\Gks|\fxcks(\omega)$ stands for $\Exc=-\frac{i}{2}\int \Gd\Sigmaxc[\Gks]$, where the $G \Wbt$ self-energy is built using the KS Green's function and KS xc kernel. The Dyson equation is then solved using this self-energy, and the resulting Green's function $\Gd$ is used in the integral. Note that while $\Gd$ is not the same in the KS and QP frameworks, we do not highlight this difference in the notation, since it is clear from the context. Comparison of the various flavors allows us to illustrate the importance of the consistency requirement advocated in Sec. \ref{xc theoretical development}. For subsequent investigation, we also indicate by $| \fxcb(\omega)$ whether the exact consistent $\fxcb(\omega)$ is used or further approximations are made, e.g., $|\fxcb(\omega=0)$.  
Fig. \ref{consistent xc} shows the difference to the exact xc energy $\Exc$. The results in the left panel were obtained using the exact consistent $\fxcb(\omega)$.
As predicted by Eq. \eqref{xc GM expression}, the two consistent calculations $\Gks|\Gks|\fxcks(\omega)$ and $\Gqp|\Gqp|\fxcqp(\omega)$ both yield the exact result. Instead, when solution of the Dyson equation is used for $G^{\text{out}}$ we obtain $\Gd|\Gks|\fxcks(\omega)$ and $\Gd|\Gqp|\fxcqp(\omega)$, which are both inconsistent and therefore not exact. The error of the former is larger than that of the latter. This can be understood, since the difference between $\Gd$ and the input Green's function is larger in the case of KS than in the case of the QP input. 
In all cases, errors are vanishing for large $t$, whereas they increase in the inconsistent calculations with decreasing $t$. Even closer to the atomic limit, all errors tend to zero. Nevertheless, the importance of consistency is nicely illustrated by this result.

\begin{figure*}[ht]
 \includegraphics[width=1\textwidth]{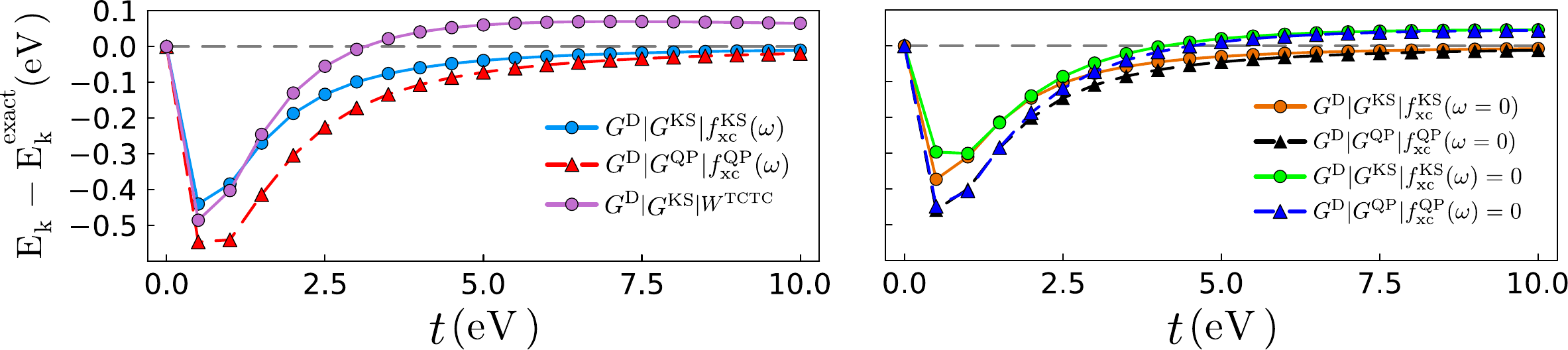}
 \caption{Kinetic energy errors as a function of the hopping parameter $t$.
 Left panel: $\Ek$ is calculated with the Green's function resulting $\Gd$ from the Dyson equation with a $G\tilde W$ self-energy (red with triangles and sky blue with dots) or with a $GW^{\rm TCTC}$ self-energy, where $W^{\rm TCTC}$ is the exact TCTC screened Coulomb interaction (violet with dots). The self-energy is built with KS ingredients (sky blue and violet) or QP ingredients (red). Right panel: $\fxc$ is approximated adiabatically (orange with dots for KS ingredients, black with triangles for QP ingredients) or completely neglected (green with dots for KS ingredients, dark blue with triangles for QP ingredients).
 \label{ek_all_plots}}
\end{figure*}

\subsubsection{Kinetic energy}

While an approximate self-energy used in the consistent scheme yields exact results for $\Exc$, no such scheme exists for the kinetic energy. Instead, by definition the result of the Dyson equation $\Gd$ is used to determine the density matrix and hence, the kinetic energy.
We will therefore examine the error introduced by various flavors of the self-energy, starting with those that can yield the exact $\Exc$.
%
%We made calculations of the kinetic energy $\text{E}_{\text{k}}$ at the $G\tilde{W}$ based on exact, adiabatic kernels and $GW$ levels,  aiming to examine the role of the $f^{\text{xc}}$ kernel and assess the impact of different approximations on the results. The 
The left panel of Fig. \ref{ek_all_plots} shows the results for $\Gd|\Gks|\fxcks(\omega)$ and $\Gd|\Gqp|\fxcqp(\omega)$. Both show errors that only vanish at large $t$ and for $t\to 0$. The KS flavor converges more quickly to the exact result with increasing $t$ than the QP version. This favors the use of the $G\tilde W$ self-energy built with KS, rather than QP, ingredients. Still, the error is significant.
%presents the outcomes of these calculations, where in this section we only discuss the results of $G\tilde{W}$ based on the exact kernels represented by the blue solid and red dashed lines with dot and triangle markers. 
However, as noted in Subsec. \ref{kinetic energy theoretical development}, with an exact interaction energy one can, in principle, also obtain the exact kinetic energy by using the virial theorem. This allows one to overcome the problem of not knowing the exact density matrix.

\begin{figure*}[ht]
 \includegraphics[width=1\textwidth]{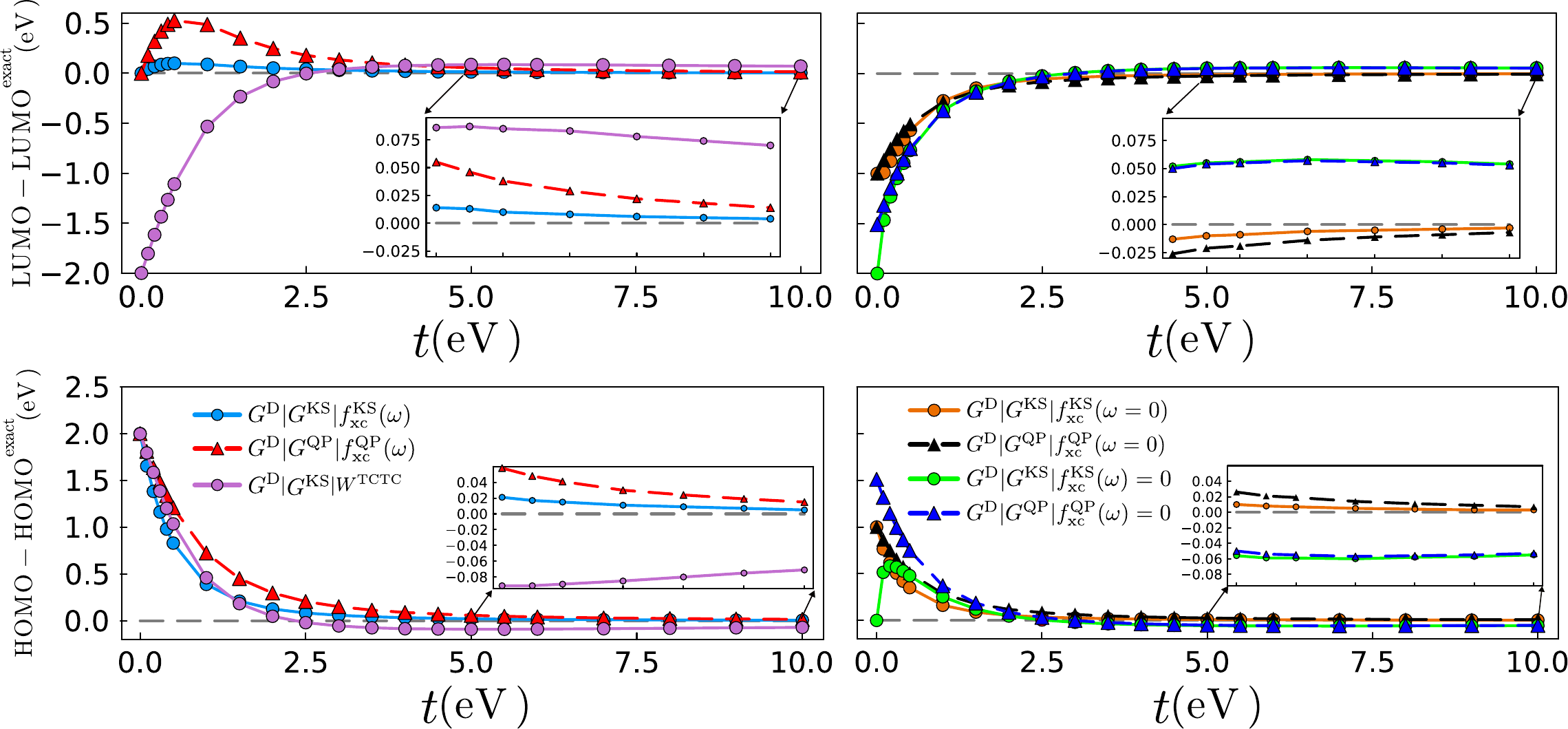}
 \caption{Error of the QP energies as a function of the hopping $t$. The LUMO and HOMO energy errors are shown in the upper and lower panels, respectively. 
 Left column: The result of the Dyson equation with a $G\tilde W$ self-energy with KS ingredients (blue with dots) or QP ingredients (red with triangles), or with a $GW$ self-energy using the exact TCTC screened Coulomb interaction (violet with dots) is shown.
Right column: The result of the Dyson equation with a $G\tilde W$ self-energy with KS ingredients using a static $\fxc(\omega=0)$ (orange with dots) or neglecting $\fxc$ (green with dots), or with QP ingredients using a static $\fxc(\omega=0)$ (black with triangles) or neglecting $\fxc$ (blue with triangles) is shown.
 \label{QP energies}}
\end{figure*}

\subsubsection{Spectra \label{spectra exact fxc}}

The situation is different for spectral properties: here, the shortcomings of an approximate Green's function cannot be overcome easily. As for the kinetic energy, the result of the Dyson equation is used to calculate the spectra. We will explore which of the flavors of the self-energy that gives an in principle exact total energy will yield the best spectral properties, and what are the remaining problems.

Let us first look at the QP peaks of the spectral function 
Fig. \ref{QP energies} shows the error of the position of the HOMO (lower panel) and of the lowest unoccupied state (LUMO) (upper panel) as a function of $t$. The two panels in the left column contain the HOMO and LUMO energy errors obtained with the exact KS or QP ingredients to build the $G\tilde W$ self-energy. While the KS and QP flavor perform very similarly for the HOMO, with small errors at larger $t$ and a significant deviation from the exact result for small $t$ that goes up to U/2 for $ t \to 0$, the LUMO is relatively well described for all $t$, and the error vanishes for $t\to 0$. Results for the LUMO are particularly satisfying when KS ingredients are used, in which case the error does not exceed 0.1 eV for any $t$. For larger $t$, above 2.5 eV, the errors become small for both HOMO and LUMO, especially in the KS flavor, where they remain well below 0.1 eV and quickly move into the meV range (see insets).  

\begin{figure*}[ht]
     \includegraphics[width=1\textwidth]{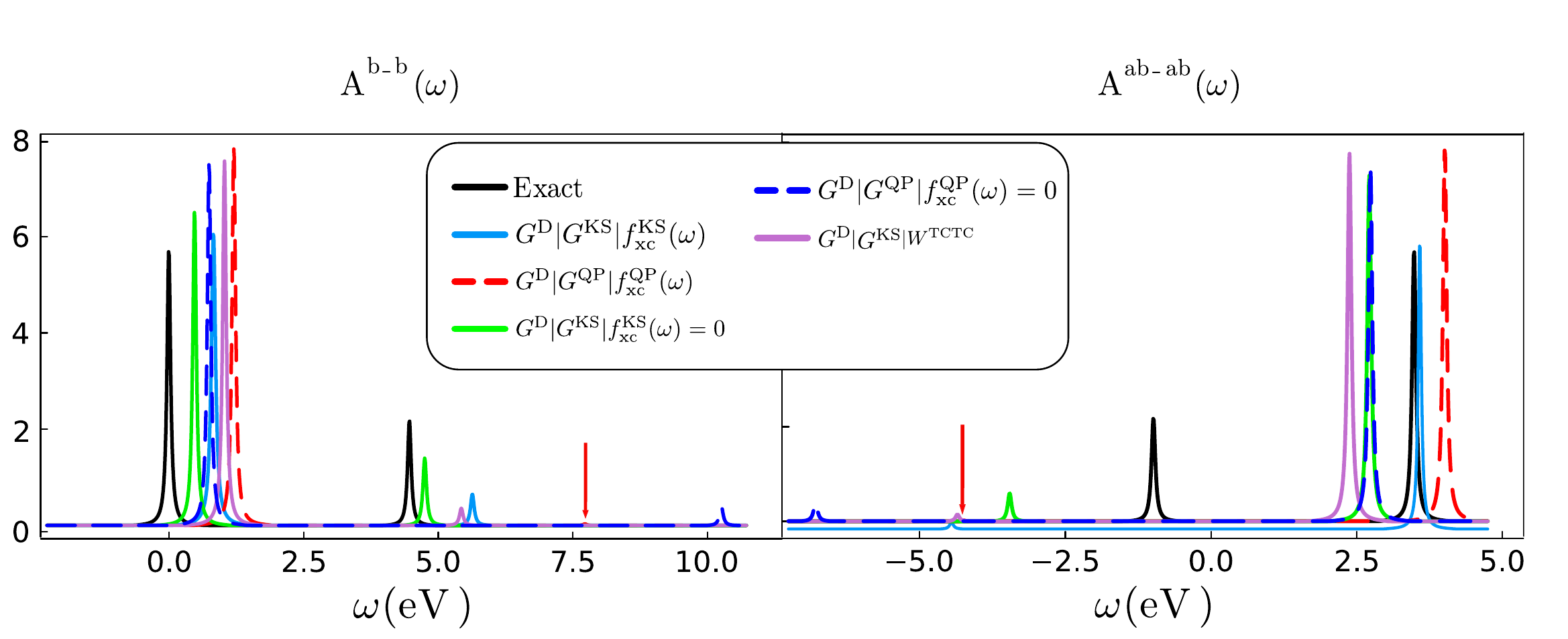}
 \caption{Bonding-bonding (left panel) and antibonding-antibonding (right panel) matrix elements of the spectral function for $U =4 $ eV and $t=0.5$ eV. The continuous black curves are the exact result. 
 The result of the Dyson equation using a $G\tilde W$ self-energy with KS ingredients and the exact $\fxcks$ is in sky blue. While the result for $\fxcks(\omega)=0$ is in green. The result of the Dyson equation using QP ingredients and the exact $\fxcqp$ is in dashed red, while the result for $\fxcqp(\omega)=0$ is in dashed blue. The red arrow indicates the position of the very weak satellite obtained when QP ingredients are used. Moreover, the result of the Dyson equation using a $GW$ self-energy with the exact TCTC screening is shown in violet. The exact HOMO is situated at 0. Note that the corresponding satellites are found at higher energies. The exact QP of the LUMO is situated at 3.5 eV, with satellites in the electron removal energy range. }
 \label{satellites} 
\end{figure*}

Beyond the QP features, Fig. \ref{satellites} shows the
entire spectral functions for $t=0.5$ eV. We will concentrate on the satellites. They are due to the peaks in the imaginary part of the self-energy, which are in turn determined by the peaks of $\tilde W$: the poles of ${\rm Im}\,\Sigmaxcb$ are situated at energies $\bar \varepsilon_i \pm \omega_j$, where $\bar \varepsilon_i $ is a removal/addition pole of the $G^{\text{in}} = \bar G$ used to build the self-energy, and $\omega_j$ is a pole of $\chi$. Not all poles are visible in all matrix elements: in the symmetric Hubbard dimer, the bonding (antibonding) matrix element of the self-energy is dominated by the addition (removal) part of $\Gin$. The bonding (antibonding) matrix element satellites are therefore found at energies higher (lower) than the LUMO (HOMO). In many real materials, all parts of the Green's function contribute to all matrix elements of the self-energy, and satellites are found on both sides of a QP. In this sense, the Hubbard dimer is an extreme case, where a given matrix element selects just one particular excitation, that may moreover not be the intuitively expected one. This does not influence our conclusions, but it is interesting to note.

The most obvious feature in Fig. \ref{satellites} is the fact that satellites are not well described in general when the exact $\fxcb (\omega)$ is used. 
Their position at $\bar \varepsilon_i \pm \omega_j$ combines two errors: the fact that the excitation energy $\omega_j$ of the $N$ electron system is used (see 
Subsec. \ref{subsec:shortcomings}), and the fact that $\bar \varepsilon_i$ can be different from the true QP energy. 
Since in our case the antibonding matrix element of the self-energy is dominated by the HOMO $\bar \varepsilon$, the exact QP energy is used in all cases studied here and the error is entirely due to the difference between the (too high) excitation energy of the $N$ electron system  with respect to the $N-1$ electron one. For the bonding matrix element, instead, the LUMO $\bar \varepsilon$ is used, which is exact when QP ingredients are used, but which is too low in the KS case. This adds to the error of the (too high) excitation energy  of the $N$ electron system with respect to the $N+1$ electron one. Since the two errors are of opposite sign, the KS ingredients yield the best result for the bonding matrix element. 
 The difference between the $N$ and $N\pm 1$ excitation energies should be of particular importance in finite systems, but an analogous error might also impact results in infinite systems with localized electrons.\cite{Ferdi2012} Note, that the problem discussed  here is different from another issue in extended systems, where the satellite position can be spoiled by the appearance of a plasmaron, a spurious solution of the QP condition that is found at some distance from the peak in the imaginary part of the self-energy \cite{Hedin1967b,Bergerse1973,Guzzo2014}. In a discrete system such as the Hubbard dimer, instead, the satellites are always found close to the position of peaks of the imaginary part of the self-energy, and the point here is that this position is calculated with the wrong number of electrons. 
 
 The biggest effect of $\fxc$ is to decrease screening, which remedies the self-screening problem for the QPs\cite{Romaniello2009,Nelson2007,Fernandez2009}, 
but which also decreases the satellite intensity because, as can be seen in Fig. \ref{approximate fxc effects on the xc energy}, $\fxc$ is always negative. Indeed, the $G\tilde W$ satellites in Fig. \ref{satellites} are of much too weak intensity. KS ingredients do a bit better than QP ones in this respect, since in this case a weaker $\fxcb$ is used (see Fig. \ref{approximate fxc effects on the xc energy}), which leads to a smaller decrease of the satellite intensities, but the result is still unsatisfactory.
This dilemma cannot be solved with such a simple vertex correction that is multipicative in frequency. 
In other words and as expected, $G\tilde W$, even with exact KS or QP ingredients, cannot yield reliable satellites.

\subsection{Impact of approximating $\fxc$}

Understanding the impact of replacing the full vertex corrected self-energy with a $G\tilde W$ one is of fundamental interest. For practical applications, one also has to face the problem that the exact $\fxc(\omega)$ is in general not known. Therefore, we also briefly examine the impact of two widely used approximations: either a complete neglect of $\fxc$, which brings us back to the $GW$ approximation with an RPA $W = W_0$, or at least an adiabatic approximation where only $\fxc(\omega=0)$ is used, since the frequency dependence of $\fxc(\omega)$ is notoriously difficult to approximate. As we will see, these approximations do not have the same impact according to the flavor (KS or QP) that is chosen, and according to the combination of ingredients. 

\subsubsection{Exchange-correlation energy: impact of approximations}

\begin{figure}[ht]
\centering
 \includegraphics[width=0.46\textwidth]{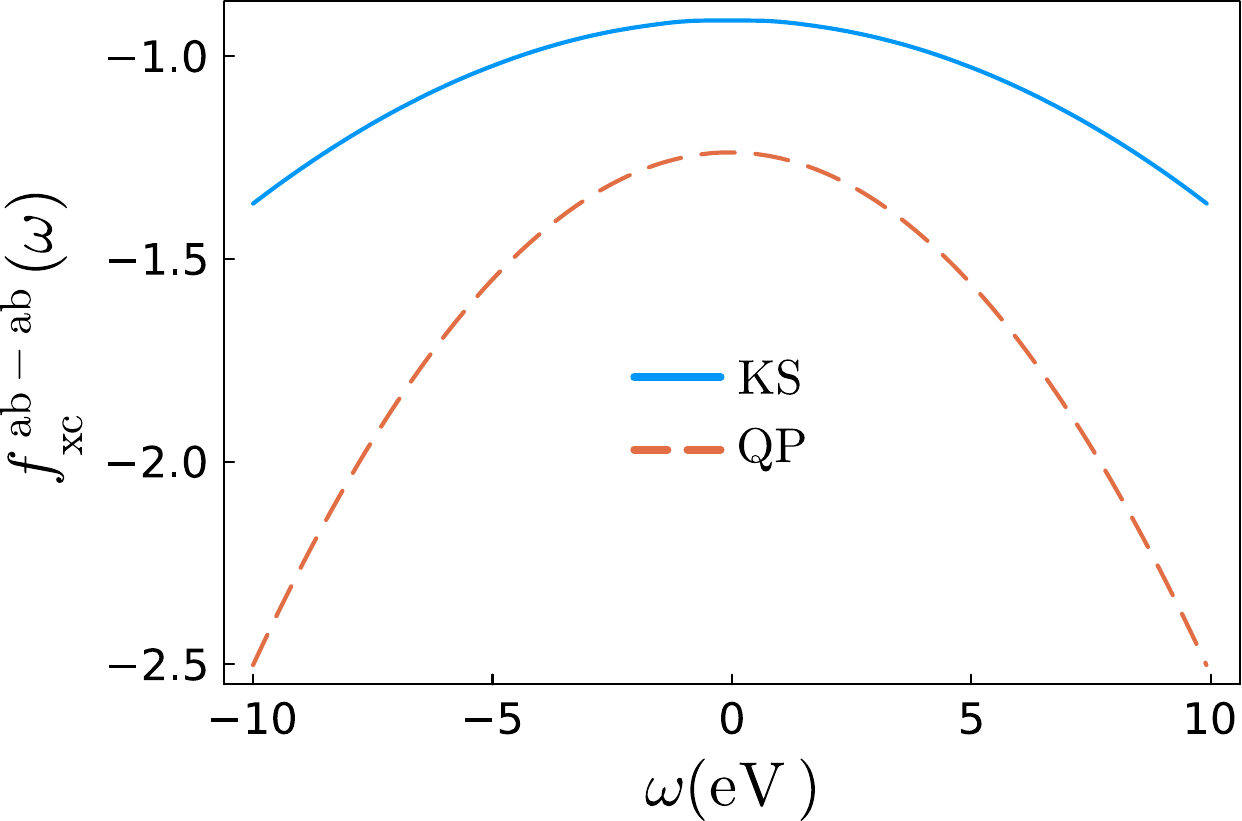}
\caption{Antibonding matrix element of $f_{\text{xc}}(\omega)$ as a function of frequency for $U=4$ eV and $t=3 $ eV. The light blue curve shows the xc kernel corresponding to the KS system, whereas the dashed orange result is the xc kernel that is consistent with QP ingredients. \label{approximate fxc effects on the xc energy}  }
\label{fxc frequency dependence}
\end{figure}

Let us first look at the quantity that is obtained exactly when $G\tilde W$ is used consistently, namely, the xc contribution $\Exc$ to the total energy. The middle panel of Fig. \ref{consistent xc} compares results using the adiabatic approximation $\fxcb(\omega=0)$ and combining the ingredients in a consistent or inconsistent way, respectively. Similarly, results in the right panel were obtained by completely neglecting $\fxc$.  In all cases, the consistent results now show an error, but it is smaller than that of the corresponding inconsistent results, which demonstrates that a consistent choice of ingredients remains essential to obtain good total energies. The impact of neglecting $\fxc$ is smaller when KS ingredients are used. The best results are obtained using the consistent KS flavor. 
When the adiabatic approximation is used, the fact that the performance of KS remains good can be explained by the fact that the quadratic frequency dependence of the kernel, which is a universal feature of $\fxc$ \cite{Botti_2005}, is milder in the KS than in the QP case, as shown in Fig. \ref{approximate fxc effects on the xc energy}.
Although approximate, the benefit of using $\fxc$ remains very important, as can be seen by comparing the middle panel and the right panel, where results on the $GW$ level with an RPA $W_0$ are given. The $GW_0$ results tend to the exact result very slowly with increasing $t$, and a part from the consistent KS flavor, they deviate significantly from the exact result in the atomic limit. The $\Gd|\Gks|f_{\text{xc}}(\omega) = 0$ flavor tends to $U/2$,while both consistent and non-consistent QP cases tend to $U/4$. 

\subsubsection{Kinetic energy: impact of approximations}

The kinetic energy is never exact in $G\tilde W$, as explained above and as illustrated in Fig. \ref{ek_all_plots}. The right panel of  Fig. \ref{ek_all_plots} also shows the impact on the kinetic energy of approximations to $\fxc$. The adiabatic approximation $\fxc(\omega=0)$ has a very moderate effect, with a tendency that is rather towards improving the results. The reason for this is the fact that the kinetic energy suffers from the underestimate of the satellite intensity discussed above, which is improved when $\fxc$ is weaker.  Neglecting the quadratic frequency-dependence of the kernel shown in Fig. \ref{approximate fxc effects on the xc energy} is therefore rather beneficial for the kinetic energy. A complete neglect of $\fxc$, instead, spoils results in the moderate to large $t$-range, while further slightly improving the small-$t$ regime, where the satellites are important. Overall, KS flavors perform slightly better than QP ones. Finally, we also show in the left panel the result of a $GW^{\text{TCTC}}$ calculation, where the self-energy is of $GW$ form using KS ingredients and the exact $\chi$ and therefore the exact $W$ is used, but where the vertex $\Gamma=1$ in the self-energy, i.e., the functional derivative in Eq. \eqref{xc self-energy} is set to 1. This means that the exact test-charge test-charge (TCTC) screening is used instead of the TCTE one that is used in the $G\tilde W$ approximation. Indeed, it would be tempting to think that a very good $W$ used in $GW$ could improve results. However, 
with respect to a standard $GW_0$ calculation using an RPA $W= W_0$, where $\fxc=0$ also in $\chi$, the results are worse, especially in the moderate to large-$t$ regime. It has been pointed out that vertex corrections in the polarizability and in the self-energy tend to cancel partially\cite{Minnhagen1974,Mahan1989}: the present result is a good illustration.

\subsubsection{Use of the adiabatic connection versus virial theorem}

Finally, we can examine the quality of the result that can be obtained by using the virial theorem, instead of approximating the kinetic energy directly, and compare to the results obtained using the AC fluctuation-dissipation theorem 
discussed in Sec. \ref{AC}. Both approaches are in principle exact, but might react differently to approximations.
   
\begin{figure}[ht]
 \includegraphics[width=0.65\textwidth]{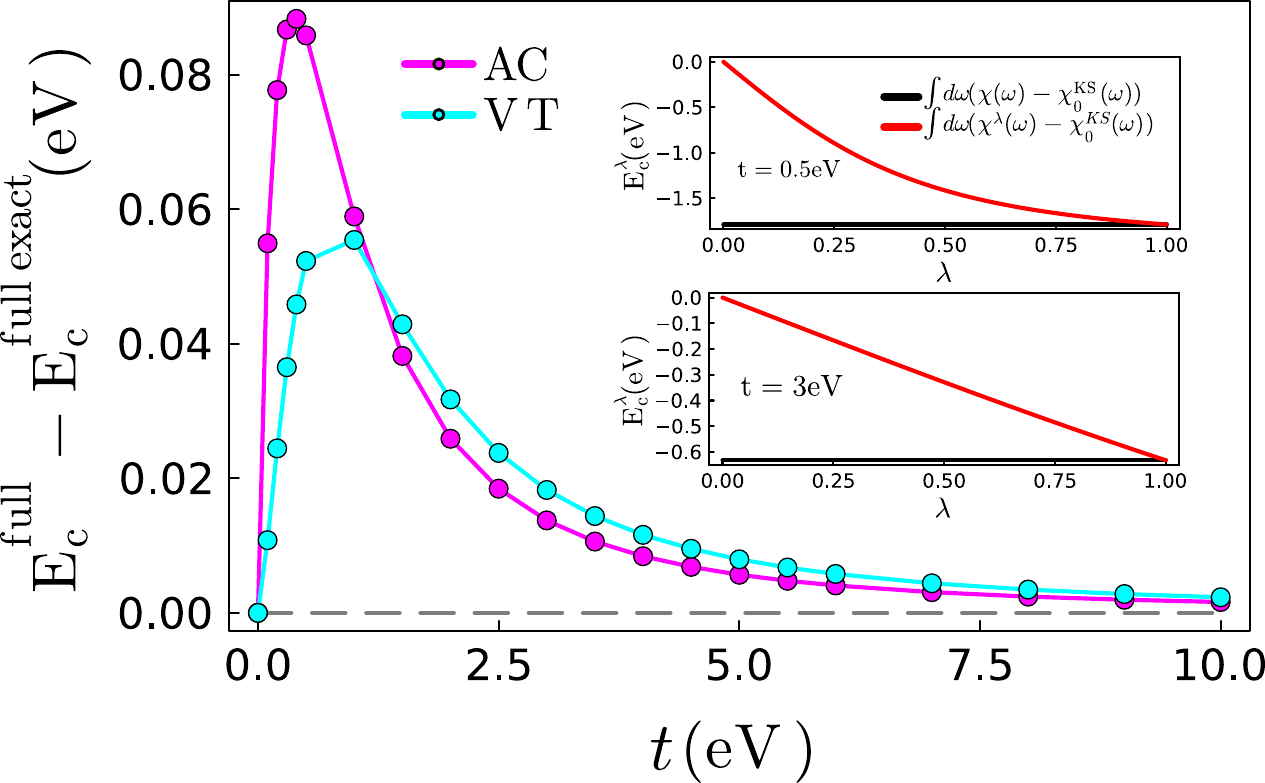}
 \caption{Error of the full correlation energy (kinetic and interaction contributions) as a function of the hopping $t$, when the adiabatic approximation $\fxc(\omega=0)$ is made and KS ingredients are used: comparison of 
 the adiabatic connection result (AC, in magenta) with the result obtained using the virial theorem (VT, in cyan). Insets: $\lambda$-resolved exact full correlation energy $\text{E}_{\text{c}}^{\text{full}} = \int_0^1{d\lambda \text{E}_{\text{c}}^{\lambda}}$, evaluated within the KS scheme, as a function of $\lambda$, for $t=0.5$ eV (upper inset) and $t=3$ eV (lower inset).
 \label{adiabatic_connection_figure}}
\end{figure}

Fig. \ref{adiabatic_connection_figure} gives the errors of the full correlation energy including interaction and kinetic contributions,  obtained using an adiabatic kernel, $\fxc(\omega=0)$ and KS ingredients. In order to use the virial theorem, one has to determine the term $S_{\rm VT}$ in  Eq. \eqref{VT}. We bypass the difficulty to adapt this equation to the Hubbard dimer by 
using the fact that here we work with the exact density in all cases, which allows us to use the exact $S_{\rm VT}$, which we obtain from the exact solution as 
$S_{\rm VT}\equiv 2\Ek+\Eint$ for all values of the hopping $t$. The resulting $S_{\rm VT}$ is then used in place of the right hand side of  Eq. \eqref{VT} in order to obtain $\Ek=(S_{\rm VT}-\Eint)/2$ for a given approximation to $\Eint$. This procedure gives the light blue curve (VT) in Fig. \ref{adiabatic_connection_figure}.
As predicted in Sec. \ref{AC}, the error is similar to the one of the AC approach using the same approximation $f_{\text{xc}}(\omega=0)$. This is indeed due to the fact that the integrand of the full correlation energy depends approximately linearly on the coupling constant $\lambda$, as one can see in the inset of Fig. \ref{adiabatic_connection_figure}. The difference of the correlation energy $\text{E}^{ \text{full} }_{\text{c}}$ is very small around $t=3$ eV where the behaviour is almost exactly linear, while the deviation is larger at the smaller $t=0.5$ eV, where a quadratic $\lambda$-dependence is clearly visible. In this small-$t$ regime, where the function is convex, the approach using the virial theorem performs better, while also avoiding the need for the $\lambda$-integration.

\subsubsection{Spectra}
The fact that $G\tilde W$ does not yield the correct spectral properties cannot be overcome, but it is still interesting to examine the effect of approximations made in practice. This is done in the right panels of Fig. \ref{QP energies} for the QP energies, and in Fig. \ref{satellites} for the satellites. For the LUMO position, both the adiabatic approximation and neglecting $\fxc$ completely lead to significant worsening of the result in the small-$t$ regime, the worst results being obtained with KS and QP $GW_0$, i.e, $\fxc = 0$, which also slightly deteriorates results at larger $t$. It is interesting to note that keeping the exact $\fxc(\omega)$ in $W$ alone, i.e., using the exact $W^{\rm TCTC} $ instead of the RPA $W = W_0$, does not fix any of these problems, as one can see in the left panel of Fig. \ref{QP energies} for the LUMO. The same is true also in the case of the HOMO. These findings are in line with observations on real systems \cite{Lewis2019}. Concerning the other approximations for the HOMO, shown in the lower right panel of Fig. \ref{QP energies}, the observation concerning the $GW$ approximation is similar to the LUMO for moderate to large $t$, whereas the adiabatic approximation is rather beneficial, especially for smaller $t$. Also a complete neglect of $\fxc$, i.e., the $GW$ solution with RPA $W =W_0$, decreases the error for small $t$, and when KS ingredients are used, the $GW_0$ results even reaches the correct $t\to 0$ limit. However, in this case the improvement is limited to a very small range of $t$ close to the atomic limit. The observed trends highlight the fact that the effect of including $\fxc$ is beneficial for the LUMO at all $t$ and for the HOMO at large $t$, but too strong  for the HOMO at small $t$. Since, as discussed in \ref{spectra exact fxc}, matrix elements of the self-energy are quite particular in the Hubbard dimer, this observation should not be generalised and further analysis will be needed to eventually turn these findings into a systematic correction, which is beyond the scope of the present work.

Finally, Fig. \ref{satellites} illustrates that including $\fxc$ in $W$ alone, i.e., performing a $GW^{\rm TCTC}$ calculation, rather worsens the satellites as compared to a $GW$ result obtained with RPA $W_0$, which illustrates again the error canceling. 
Therefore, in Fig. \ref{satellites} the best satellite results are obtained using the $GW_0$ with KS ingredients and a complete neglect of $\fxcb$. Note, however, that this is not a general finding for all values of $U/t$.

%% file: conclusion.tex
\section{Conclusions \label{conclusion}}
In conclusion, the \emph{exact} exchange-correlation contribution to the total interaction energy can be calculated using an \emph{approximate} self-energy of the form $G\tilde{W}$. Here, $\tilde W$ is a test-charge test-electron screened Coulomb interaction, which replaces the RPA or the TCTC screened interaction that are commonly used in the $GW$ approximation. Different choices for $\tilde W$ are possible, one of them being the traditionally used KS scheme, which adds an xc kernel $f_{\rm xc}(\omega)$ from linear response TDDFT to the bare Coulomb interaction in the dielectric function. For all choices the
condition is that the \GF\ and xc kernel used to build the self-energy are consistent and yield the correct density. On top of the KS choice, we have examined the case where the \GF\ is built with the exact QP energies. For all possible choices, it holds that  
the exact xc energy is obtained by integrating the approximate self-energy with the very same \GF\ that was used to build it. 
Instead, when the approximate self-energy is used in a Dyson equation and integrated with the resulting \GF , the results carry an error. The importance of consistency between the \GF\ used to build the self-energy and the \GF\ used for the integration may explain the success of self-consistent $GW$ total energy calculations, which indeed fulfill the requirement that the self-energy is integrated with the \GF\ that is used to build it. Here, we show that one can obtain good quality results by being consistent without carrying out self-consistent calculations.

The exact correlation contribution to the kinetic energy cannot be accessed in the same way. Instead, we propose to use the virial theorem. We have studied the impact of widely used approximations to this approach, and compared with the use of the adiabatic connection fluctuation dissipation theorem. Our general derivation predicts that the final errors are similar, without the need of a coupling constant integration in the present approach. 

Using the approximate self-energies in the Dyson equation leads to approximate \GF s and therefore, to approximate spectral functions. Still, $G\tilde W$ yields overall better QPs than $GW$, and since the computational difficulty is similar, it should be preferred. The satellite problem, instead, cannot be fixed in this way.

All statements have been illustrated for the symmetric half-filled Hubbard dimer, confirming our conjectures and highlighting the fact that results obtained using KS ingredients are overall superior and less impacted by additional approximations with respect to results obtained using QP ingredients. While the Hubbard dimer is a simple model, our findings relie on derivations that are valid for the general case, and they should open the way for interesting applications to more realistic systems. 
%In this paper, we have presented a novel approach to obtain the exact xc energy without relying on the exact xc self-energy or the exact 1-GF. Instead, we have demonstrated that using an xc self-energy based on a two-point $f^{\text{xc}}$ kernel is sufficient to reproduce the exact xc energy, as long as the ingredients used in the calculation are consistent. This consistency requirement allows for the use of different kernels, as long as the chosen $\chi_0$ contains the exact density.

\begin{acknowledgement}
The authors acknowledge the fruitful discussions with Steffen Backes, Fabien Bruneval, Kieron Burke and Steven Crisostomo. \\
This project has received funding from the European Union's Horizon 2020 research and innovation programme under grant agreement No 800945 — NUMERICS — H2020-MSCA-COFUND-2017.
\end{acknowledgement}

%% file: appendices.tex
\begin{suppinfo}
\setcounter{section}{0}
The exact and $GW$ solutions for the symmetric Hubbard dimer model at half-filling (two electrons) are given in literature \cite{Romaniello2009}.
In this section, we provide the solutions for the model within the $GW$ and $G\tilde{W}$ approximations using both the KS and QP flavors.

\section{$GW$ solutions \label{GW solutions}}

The two ingredients needed to calculate $\Sigmaxc$ at the $GW$ level are the \GF\ and the screened Coulomb interaction.

\begin{enumerate}
 \item  The exact Kohn-Sham \GF\ and the exact Quasi-Particle (QP) \GF\ 
in the dimer sites basis read respectively
 \begin{equation}
  \Gks_{ij\sigma}(\omega) = \frac{1}{2} \bigg( \frac{1}{\omega-(\epsilon_0+t - (c-U)/2)-i\eta } + \frac{(-1)^{i-j}}{\omega-(\epsilon_0+3t-(c-U)/2)+i\eta} \bigg),
 \end{equation}
 \begin{equation}
  \Gqp_{ij\sigma}(\omega) = \frac{1}{2} \bigg( \frac{1}{\omega-(\epsilon_0+t - (c-U)/2 ) -i\eta } + \frac{(-1)^{i-j}}{\omega-(\epsilon_0-t+(c+U)/2)+i\eta} \bigg),
 \end{equation}
 where $c = \sqrt{16 t^2 + U^2}$. The KS \GF\ ($\Gks$) is obtained by introducing an energy shift to the poles of the 
 non-interacting \GF\, such that HOMO energy becomes exact\cite{Perdew1982a}. The QP GF $(\Gqp$) equals the 
exact GF without the satellite contributions and with the quasiparticle intensities set to $1$.
\item We use both the exact screened Coulomb interaction $W$ and approximations denoted $\Woks$ or $\Woqp$, 
depending on the choice of the \GF\ used to compute the polarizability. The $\Woks$ and $\Woqp$ are calculated within 
the Random Phase Approximation (RPA), using the following irreducible polarizabilities
\begin{equation}
 \Prpaks(1,2) = -i \Gks(1,2^+)\Gks(2,1^+), 
\end{equation}
and
\begin{equation}
\Prpaqp(1,2) = -i \Gqp(1,2^+)\Gqp(2,1^+), 
\end{equation}
respectively for the KS and QP cases. They have the following analytical expressions
\begin{equation}
\Prpaksijs(\omega) = \frac{(-1)^{i-j}}{4} \bigg( \frac{1}{\omega-2t+i\eta}-\frac{1}{\omega+2t-i\eta} \bigg), 
\label{prpaks}
\end{equation}
\begin{equation}
\Prpaqpijs(\omega) = \frac{(-1)^{i-j}}{4} \bigg( \frac{1}{\omega+(2t-c)+i\eta}-\frac{1}{\omega-(2t-c)-i\eta} \bigg). 
\label{prpaqp}
\end{equation}
For the exact $W$, we use the exact reducible polarizability $\chi$, which is related to the 2-particle Green's Function ($G_2$)
\begin{equation}
\chi(1,2)=-i G(1,1^+)G(2,2^+) +i G_2(1,2,1^+,2^+),  
\end{equation}
where $G_2$ is the 2-GF. In the Hubbard dimer site ($ij$) basis we have
\begin{multline}
 \chi_{ij\sigma_1\sigma_2}(\omega ) =  \sum_{s\neq0}  \bigg[\braket{N_0|\hat{c}^{\dagger}_{i\sigma_1} \hat{c}_{i\sigma_1} \ket{N_s} \bra{N_s}  \hat{c}^{\dagger}_{j\sigma_2} \hat{c}_{j\sigma_2}|N_0} \frac{1}{\omega + ( \text{E}^N_0-\text{E}^N_s)+i\eta}  \\ - \braket{N_0|\hat{c}^{\dagger}_{j\sigma_2} \hat{c}_{j\sigma_2} \ket{N_s} \bra{N_s}  \hat{c}^{\dagger}_{i\sigma_1} \hat{c}_{i\sigma_1}|N_0} \frac{1}{\omega- (\text{E}^N_0-\text{E}^N_s)-i\eta} \bigg]  \, ,
\end{multline}
which leads to the following solutions
\begin{multline}
\chi_{ij\uparrow\uparrow}(\omega)=  \frac{(-1)^{i-j}}{2a^2} \bigg(  \frac{1}{\omega-(c+U)/2 +i\eta} -\frac{1}{\omega+(c+U)/2 -i\eta}  \bigg) \\ 
+ (-1)^{i-j}\frac{16t^2}{2a^2(c-U)^2} \bigg(\frac{1}{\omega-(c-U)/2 +i\eta} - \frac{1}{\omega+(c-U)/2 -i\eta}\bigg), 
\end{multline}
\begin{multline}
\chi_{ij\uparrow\downarrow}(\omega) =  \frac{(-1)^{i-j}}{2a^2} \bigg(  \frac{1}{\omega-(c+U)/2 +i\eta} - \frac{1}{\omega+(c+U)/2 -i\eta}  \bigg) \\ 
- (-1)^{i-j}\frac{16t^2}{2a^2(c-U)^2} \bigg(\frac{1}{\omega-(c-U)/2 +i\eta} - \frac{1}{\omega+(c-U)/2 -i\eta}\bigg),
\end{multline} 
where $a^2 = 2\bigg( \frac{16t^2}{(c-U)^2} + 1 \bigg)$ and $\chi_{ij \uparrow\uparrow}=\chi_{ij \downarrow\downarrow}$, 
$\chi_{ij\uparrow\downarrow} = \chi_{ij\downarrow\uparrow}$.

The spin-independent $\chi$ matrix in the site basis is a sum over spins, i.e. 
$\chi_{ij}(\omega) = \chi_{ij\uparrow\uparrow} + \chi_{ij\uparrow\downarrow} + \chi_{ij\downarrow\uparrow} + \chi_{ij\downarrow\downarrow}$. In the bonding and anti-bonding (b/ab) basis, $\chi$ reads

\begin{equation}
    \begin{pmatrix}
0 & 0 \\
0& \chi_{\text{ab-ab}}(\omega)
    \end{pmatrix},
    \label{exact chi in the b and ab basis}
\end{equation}
where $\chi_{\text{ab-ab}}(\omega) = 2(\chi_{11}(\omega) + \chi_{22}(\omega))$.
We can now write the screened Coulomb interaction $W$ in the different flavors. The exact screened Coulomb interaction is
\begin{align*}
 W(1,2) &= \vcoul(1,2) + \int{d(34) \vcoul(1,3)P(3,4)W(4,2)}, \\ 
        &= \vcoul(1,2) + \int{d(34) \vcoul(1,3)\chi(3,4)\vcoul(4,2)}\,, 
\end{align*}
or, in the site basis
\begin{align*}
W_{ij}(\omega) &= U\delta_{ij} + U \sum_{k\sigma} P_{ik\sigma}(\omega)W_{kj}(\omega), \\
               &= U\delta_{ij} + U^2\sum_{\sigma\sigma'=\uparrow,\downarrow} \chi_{ij\sigma\sigma'}(\omega)\,, 
\end{align*}
which leads to 
\begin{equation*}
 W_{ij}(\omega)= U\delta_{ij} + (-1)^{i-j}\frac{2U^2}{a^2} \bigg(  \frac{1}{\omega-(c+U)/2 +i\eta}- \frac{1}{\omega+(c+U)/2 -i\eta}  \bigg)\,.
\end{equation*}
By using Eq.s \eqref{prpaks}\eqref{prpaqp}, we find
\begin{equation}
 \Woksij(\omega)= U\delta_{ij} + (-1)^{i-j}\frac{U^2 t}{h} \bigg(  \frac{1}{\omega- h + i\eta}- \frac{1}{\omega+ h -i\eta}  \bigg)\,,
\end{equation}
where $h = \sqrt{4t^2 + 4Ut}$, and 
\begin{equation}
 \Woqpij(\omega)= U\delta_{ij} + (-1)^{i-j}\frac{U^2 (c/2 - t)}{h'} \bigg(  \frac{1}{\omega- h' + i\eta}- \frac{1}{\omega+ h' -i\eta}  \bigg)\,,
\end{equation}
where $h' = \sqrt{ (2t- c)^2 + 4U(c/2-t)}$. Now, by using $\Gks$, $\Gqp$, $\Woks$, $\Woqp$ and $W$ we calculate the different 
flavors of $\Sigmaxc$ by integrating in frequency space. 
Finally, we convert $\Sigmaxc$ in Eq. \eqref{Wtcte} to frequency space. Note that the use of multiple infinitesimals in Eq. \eqref{Wtcte} is not important for the self-energy itself, but for the calculation of $\Exc$, as the respective weight of the infinitesimals  in the different contributions indicates the contour that is to be used in the integral. So
\begin{equation}
\Sigmaxcijs(\omega) = \frac{i}{2\pi}\int{d\omega'\, G_{0,ij\sigma}(\omega'+\omega) W_{0,ji}(\omega') e^{3i\omega'\eta}}\,,
\end{equation}
where $G_0$ can be $\Gks$ or $\Gqp$, and $W_0$ can be $\Woks$, $\Woqp$ or $W$. 
\end{enumerate}

The solutions of the different $GW$ flavors are
\begin{multline}
\Sigma^{\Gks\Woks}_{ \text{xc}, ij\sigma}(\omega) = -\frac{U}{2} \delta_{ij} + \frac{U^2t}{2h} \bigg(  \frac{1}{\omega - (\epsilon_0+3t - (c-U)/2 +h)+i\eta} \\  + \frac{ (-1)^{i-j} e^{-3i\omega\eta }}{\omega - (\epsilon_0+t-(c-U)/2-h)-i\eta} \bigg)\, ,
\end{multline}

\begin{equation}
\Sigma^{\Gks W}_{\text{xc}, ij\sigma}(\omega) = -\frac{U}{2} \delta_{ij} + \frac{U^2}{a^2} \bigg( \frac{1}{\omega - (\epsilon_0+3t+U)+i\eta} + \frac{(-1)^{i-j} e^{-3i\omega\eta} }{\omega - (\epsilon_0+t-c)-i\eta} \bigg)\,,
\end{equation}

\begin{multline}
\Sigma_{ \text{xc},ij\sigma}^{\Gqp\Woqp} = -\frac{U}{2}\delta_{ij} + \frac{U^2(\frac{c}{2} -t)}{2h'}\bigg( \frac{1}{\omega-(\epsilon_0-t+(c+U)/2 + h')+i\eta} +  \\ \frac{(-1)^{i-j} e^{-3i\omega\eta}}{\omega-(\epsilon_0+t-(c-U)/2 -h')-i\eta}\bigg)\,,
\end{multline}
and, 
\begin{equation}
\Sigma_{ \text{xc}, ij\sigma}^{\Gqp W } = -\frac{U}{2}\delta_{ij} + \frac{U^2 }{a^2}\bigg( \frac{1}{\omega-(\epsilon_0-t+c+U)+i\eta}  + \frac{(-1)^{i-j} e^{-3i\omega\eta}}{\omega-(\epsilon_0+t-c)-i\eta}\bigg). 
\end{equation}

The non-interacting $\chi_0(1,2) = -i G(1,2^+)G(2,1^+)$ does not have the same structure 
as the exact $\chi$ for the Hubbard dimer, when $G$ is the exact \GF\ . In fact,
\begin{multline}
\chi^{GG}_{0, ij}(\omega) = (-1)^{i-j}\frac{\big(1 + \frac{4t}{c-U}\big)^4 }{2a^4} \times \bigg( \frac{1}{\omega+ 2t -c +2i\eta} - \frac{1}{\omega - 2t +c -2i\eta}\bigg)  + \\ \frac{ \big(1 + \frac{4t}{c-U}\big)^2 \big( 1 - \frac{4t}{c-U}\big)^2    }{a^4} \times \bigg( \frac{1}{\omega-c +2i\eta} - \frac{1}{\omega+ c- 2i\eta} \bigg)  \\ + (-1)^{i-j}\frac{\big(1 - \frac{4t}{c-U}\big)^4 }{2a^4} \times \bigg( \frac{1}{\omega- 2t -c +2i\eta} - \frac{1}{\omega + 2t +c -2i\eta}\bigg) \,,
\label{chi0 GG}
\end{multline}
which yields, in the b/a-b basis
\begin{equation}
\chi_0^{GG}(\omega) = \begin{pmatrix}
    2 C_2 & 0 \\
    0 & 2C_1 + 2C_3
    \end{pmatrix}\,,
\end{equation}
where $C_1$, $C_2$ and $C_3$ correspond to the first, second and last  term in Eq. \ref{chi0 GG}, respectively. The fact that the bonding-bonding matrix element does not vanish, contratry to the exact interacting $\chi$ Eq. \ref{exact chi in the b and ab basis}, explains why no $\fxc$ can be found that would link $\chi_0^{GG}$ and $\chi$ in a Dyson equation. 
\section{$G\tilde{W}$ solutions \label{appendix beyond GW}}

The test-charge test-electron screened interaction is defined as

\begin{equation}
\tilde{W}(1,2) = \vcoul(1,2) + \int{d(34)}\bigg( \vcoul(1,3) \\ + \fxc(1,3)\bigg)\chi(4,2)\vcoul(4,2) \,.
\end{equation}

The two $\fxc$ kernels that we used in the main text are given by the matrix equations below
\begin{equation}
\fxcks(\omega) = \left[\chioks(\omega)\right]^{-1} - \left[\chi(\omega)\right]^{-1} - \vcoul \,,
\label{fxcmatrixequation}
\end{equation}
\begin{equation}
\fxcqp(\omega) = \left[\chioqp(\omega)\right]^{-1} - \left[\chi(\omega)\right]^{-1} - \vcoul \,.
\end{equation}
In the bonding-antibonding basis, because of Eq. \eqref{exact chi in the b and ab basis} the KS $\chi_0$ cannot be inverted and $\fxc^{b-b}$ is not determined. Instead, $\fxc^{\text{ab-ab}}=\frac{1}{2\chi_{0,11}}-\frac{1}{2\chi_{11}}-U$,
where $\chi^{\mathrm{KS}}_{0,11}(\omega) = \chi^{\mathrm{KS}}_{0,11\uparrow}(\omega) +\chi^{\mathrm{KS}}_{0,11\downarrow}(\omega) $ and $\chi_{11}(\omega) = \chi_{11\uparrow\uparrow}(\omega)+\chi_{11\uparrow\downarrow}(\omega)+\chi_{11\downarrow\uparrow}(\omega)+\chi_{11\downarrow\downarrow}(\omega) $. $f_{\text{xc}}^{\text{b-b}}$ and $f_{\text{xc}}^{\text{ab-ab}}$ are the bonding-bonding and antibonding-antibonding elements of the $f_{\text{xc}}$ matrix. This leads to
\begin{equation}
f_{\text{xc}}^{\mathrm{KS,ab-ab}}(\omega) = \omega^2 \bigg(\frac{1}{4t} -\frac{a^2}{4(c+U)} \bigg) - t  + \frac{a^2(c+U)}{16} - U \, ,
\label{fxc abab KS}
\end{equation}
and
\begin{equation}
f_{\text{xc}}^{\mathrm{QP,ab-ab}}(\omega) = \omega^2\bigg( \frac{1}{2c-4t} - \frac{a^2}{4(c+U)} \bigg) + 2t-c  + \frac{a^2 (c+U)}{16} - U \,,
\label{fxc abab QP}
\end{equation}
respectively for the KS and QP cases. In the symmetric Hubbard dimer, $\fxc$ does not have poles. Its frequency dependence is quadratic. Comparison of Eq. \eqref{fxc abab KS} with Eq. \eqref{fxc abab QP} shows that, since $\frac{1}{2c-4t} - \frac{a^2}{4(c+U)} >\frac{1}{4t} - \frac{a^2}{4(c+U)}$,  $\fxcqp$ varies more strongly than $\fxcks$ with $\omega$, as it is also shown in Fig. \ref{fxc frequency dependence}. 

The self-energies based on these kernels read
\begin{multline}
\Sigma_{ \text{xc}, ij\sigma}(\omega)=i\int{\frac{d\omega_1}{2\pi} G_{ij\sigma}(\omega_1+\omega)W_{ji}(\omega_1)e^{3i\omega_1\eta} } \\+ iU\sum_{m=1,2}\int{\frac{d\omega_1}{2\pi} G_{ij\sigma}(\omega_1+\omega) f_{xcjm}(\omega_1) \chi_{mi}(\omega_1)e^{3i\omega_1 \eta}} \,,
\end{multline}
which yields  
\begin{multline}
\Sigma^{\Gks\tilde{W}^{\mathrm{KS}} }_{ \text{xc},ij\sigma}(\omega) = -\frac{U}{2}\delta_{ij} + (-1)^{i-j}\bigg( \frac{U }{4}  - \frac{U(c+U)}{4ta^2}\bigg) e^{-3i\omega\eta}  + \\ \frac{U}{4a^2}\times \frac{(c+U)^2/4 - 4t^2 }{t} \bigg( \frac{1}{ \omega-(\epsilon_0 + 3t +U)+i\eta } + \frac{(-1)^{i-j} e^{-3i\omega\eta}}{\omega-(\epsilon_0+t-c)-i\eta}\bigg)\,,
\end{multline}
\begin{multline}
 \Sigma^{\Gqp\tilde{W}^{\mathrm{QP}} }_{ \text{xc},ij\sigma}(\omega) = -\frac{U}{2}\delta_{ij} + (-1)^{i-j} \bigg( \frac{U }{4}  - \frac{U(c+U)}{a^2 (2c-4t)} \bigg)e^{-3i\omega\eta} \\
 + \frac{U}{a^2}\times \frac{(c+U)^2/4 - (2t-c)^2 }{(2c-4t)} \bigg( \frac{1}{\omega+(\epsilon_0-t+c+U)+i\eta} + \frac{(-1)^{i-j} e^{-3i\omega\eta}}{\omega-(\epsilon_0+t-c)-i\eta}\bigg)\, ,
\end{multline}
in which $\tilde{W}$ is the TCTE screened interaction based on the exact $\chi$ and  consistent $\fxc$ kernel.

When we use the the adiabatic approximation for $\fxc$ to evaluate $\chi$ and $\Sigmaxc$, the Dyson equation becomes
\begin{equation}
 \chiad(\omega) = \bigg( \chi_0^{-1}(\omega) - f_{ \text{xc} }(\omega=0) -\vcoul \bigg)^{-1}\,.  
 \label{adiabatic chi}
\end{equation}
 We calculate two different $\chiad$ depending on the choice of $\chi_0$ and corresponding $\fxc(\omega=0)$. So, we have $\chiksad$ and $\chiqpad$ when $\chioks$, $\fxcks(\omega=0)$ and $\chioqp$, $\fxcqp(\omega=0)$ are used respectively in the equation above. This yields 
\begin{multline}
 \Sigma^{\Gks\tilde{W}^{\mathrm{KS}}_{\text{adiab}}}_{\text{xc},ij\sigma}(\omega=0) = -\frac{U}{2}\delta_{ij} +\\ \frac{Ut}{2\omega_1}\bigg( \frac{1}{\omega-(\epsilon_0+3t-(c-U)/2 +\omega_1)+i\eta } + \frac{(-1)^{i-j} e^{-3i\omega\eta}}{\omega-(\epsilon_0+t-(c-U/2)-\omega_1)-i\eta}\bigg)\,, 
\end{multline}
where $\omega_1 = \sqrt{4t^2 +2t f_{\text{Hxc}}^{\mathrm{KS}}(\omega=0) } $, where $f_{\text{Hxc}}^{\mathrm{KS}}(\omega=0) =  f_{\text{xc},11}^{\mathrm{KS}}(\omega) -  f_{\text{xc},12}^{\mathrm{KS}}(\omega=0) +2U = -2t + \frac{a^2(c+U)}{8}$
For the QP ingredients, we have, similarly
\begin{multline}
 \Sigma^{\Gqp\tilde{W}^{\mathrm{QP}}_{\text{adiab}}}_{ \text{xc}, ij\sigma}(\omega=0) =-\frac{U}{2}\delta_{ij} + \frac{U(2c-4t)}{8\omega_2} \\
\times \bigg( \frac{1}{\omega-(\epsilon_0-t+(c+U)/2 +\omega_2)+i\eta } + \frac{(-1)^{i-j}e^{-3i\omega\eta}}{\omega-(\epsilon_0+t-(c-U/2)-\omega_2)-i\eta}\bigg)\,, 
\end{multline}
where $\omega_2 = \sqrt{ (2t-c)^2 + (2c-4t)f_{\text{Hxc}}^{\mathrm{QP}}(\omega=0)  }$, with $f_{\text{Hxc}}^{\mathrm{QP}}(\omega=0) = \frac{2t-c}{2} + \frac{a^2(c+U)}{16}$. Note that $\tilde{W}_{\text{adiab}}$ is the TCTE screened interaction that includes $f_{\text{xc}}(\omega=0)$ and $\chiad(\omega)$ within the two different schemes.

\section{Total energy contributions for the Hubbard dimer}
The xc and kinetic energy contributions to the total energy, given in the Galitskii-Migdal formula in Eq. \eqref{GM equation} are written in the site basis and frequency space of the Hubbard dimer respectively as follows

\begin{equation}
   \text{E}_{\text{xc}} = - \frac{i}{2}\sum_{ij\sigma} \int_{-\infty}^{+\infty}{\frac{d\omega}{2\pi} \, \Sigma_{\text{xc},ij\sigma}(\omega) G_{ij\sigma}(\omega) e^{2i\omega\eta} } \, , 
   \label{exc in the site basis}
\end{equation}

\begin{equation}
   \text{E}_{\text{k}} = i t \sum_{ij,i\neq j,\sigma} \int_{-\infty}^{+\infty}{ \frac{d\omega}{2\pi} \, G_{ij\sigma}(\omega) e^{i\omega \eta}  }\,. 
   \label{ek in the site basis}
\end{equation}

\section{Computational details}
The entire framework for this work has been developed using an in-house code, using the Julia programming language.\cite{Julia2017}
For the purpose of performing the energy integrals, we use 
the `quadgk' library, that relies on Gauss-Kronod 
quadratures.\cite{Laurie1997} Additionally, to visualize our 
findings effectively, we rely on the `Plots.jl' library,\cite{PlotsJL} 
coupled with the \href{https://github.com/jheinen/GR.jl}{GR backend}.
While we provide the analytic solutions and equations for the time-ordered quantities, the numerical calculations have been performed using the retarded Green's function framework,\cite{Spataru2004,Honet2022} which yields numerically stable results for small $t$. Retarded $G^R$ and $\Sigma^R$ are obtained from the above equations with the usual sign changes of the imaginary infinitesimals. The total energy contributions defined in Eq. \eqref{exc in the site basis} and Eq. \eqref{ek in the site basis} become

\begin{equation}
   \text{E}_{\text{xc}} = - \frac{1}{2\pi}\sum_{ij\sigma} \int_{-\infty}^{\mu}{d\omega \,\text{Im} (\Sigma^R_{\text{xc},ij\sigma}(\omega) G^R_{ij\sigma}(\omega))  } \, , 
   \label{exc retarded in the site basis}
\end{equation}

\begin{equation}
   \text{E}_{\text{k}} =  \frac{t}{\pi}  \sum_{ij,i\neq j,\sigma} \int_{-\infty}^{\mu}{ d\omega \, \text{Im} G^R_{ij\sigma}(\omega)   }\,,
   \label{ek in the site basis}
\end{equation}
where $\mu$ is the chemical potential.\\

The code of this project, called ``Symmetric Hubbard Dimer'', is available at the following address: \href{https://gitlab.com/tsg1860938/symmetric-hubbard-dimer}{https://gitlab.com/tsg1860938/symmetric-hubbard-dimer}

\end{suppinfo}

%% file: main.bbl
\providecommand{\latin}[1]{#1}
\makeatletter
\providecommand{\doi}
  {\begingroup\let\do\@makeother\dospecials
  \catcode`\{=1 \catcode`\}=2 \doi@aux}
\providecommand{\doi@aux}[1]{\endgroup\texttt{#1}}
\makeatother
\providecommand*\mcitethebibliography{\thebibliography}
\csname @ifundefined\endcsname{endmcitethebibliography}
  {\let\endmcitethebibliography\endthebibliography}{}
\begin{mcitethebibliography}{97}
\providecommand*\natexlab[1]{#1}
\providecommand*\mciteSetBstSublistMode[1]{}
\providecommand*\mciteSetBstMaxWidthForm[2]{}
\providecommand*\mciteBstWouldAddEndPuncttrue
  {\def\EndOfBibitem{\unskip.}}
\providecommand*\mciteBstWouldAddEndPunctfalse
  {\let\EndOfBibitem\relax}
\providecommand*\mciteSetBstMidEndSepPunct[3]{}
\providecommand*\mciteSetBstSublistLabelBeginEnd[3]{}
\providecommand*\EndOfBibitem{}
\mciteSetBstSublistMode{f}
\mciteSetBstMaxWidthForm{subitem}{(\alph{mcitesubitemcount})}
\mciteSetBstSublistLabelBeginEnd
  {\mcitemaxwidthsubitemform\space}
  {\relax}
  {\relax}

\bibitem[Hohenberg and Kohn(1964)Hohenberg, and Kohn]{Hohenberg1964}
Hohenberg,~P.; Kohn,~W. Inhomogeneous Electron Gas. \emph{Phys. Rev.}
  \textbf{1964}, \emph{136}, B864--B871\relax
\mciteBstWouldAddEndPuncttrue
\mciteSetBstMidEndSepPunct{\mcitedefaultmidpunct}
{\mcitedefaultendpunct}{\mcitedefaultseppunct}\relax
\EndOfBibitem
\bibitem[Gilbert(1975)]{Gilbert1975}
Gilbert,~T.~L. Hohenberg-Kohn theorem for nonlocal external potentials.
  \emph{Phys. Rev. B} \textbf{1975}, \emph{12}, 2111--2120\relax
\mciteBstWouldAddEndPuncttrue
\mciteSetBstMidEndSepPunct{\mcitedefaultmidpunct}
{\mcitedefaultendpunct}{\mcitedefaultseppunct}\relax
\EndOfBibitem
\bibitem[Donnelly and Parr(2008)Donnelly, and Parr]{Donnelly2008}
Donnelly,~R.~A.; Parr,~R.~G. {Elementary properties of an energy functional of
  the first‐order reduced density matrix}. \emph{The Journal of Chemical
  Physics} \textbf{2008}, \emph{69}, 4431--4439\relax
\mciteBstWouldAddEndPuncttrue
\mciteSetBstMidEndSepPunct{\mcitedefaultmidpunct}
{\mcitedefaultendpunct}{\mcitedefaultseppunct}\relax
\EndOfBibitem
\bibitem[Levy(1979)]{Levy1979}
Levy,~M. Universal variational functionals of electron densities, first-order
  density matrices, and natural spin-orbitals and solution of the
  <i>v</i>-representability problem. \emph{Proceedings of the National Academy
  of Sciences} \textbf{1979}, \emph{76}, 6062--6065\relax
\mciteBstWouldAddEndPuncttrue
\mciteSetBstMidEndSepPunct{\mcitedefaultmidpunct}
{\mcitedefaultendpunct}{\mcitedefaultseppunct}\relax
\EndOfBibitem
\bibitem[Martin \latin{et~al.}(2016)Martin, Reining, and Ceperley]{Martin2016}
Martin,~R.; Reining,~L.; Ceperley,~D. \emph{Interacting Electrons: Theory and
  Computational Approaches}; Cambridge University Press, 2016\relax
\mciteBstWouldAddEndPuncttrue
\mciteSetBstMidEndSepPunct{\mcitedefaultmidpunct}
{\mcitedefaultendpunct}{\mcitedefaultseppunct}\relax
\EndOfBibitem
\bibitem[V.M.~Galitskii(1950)]{Galitskii1950}
V.M.~Galitskii,~A.~M. Application of Quantum Field Theory Methods to the Many
  Body Problem. \emph{JETP} \textbf{1950}, \emph{7}, 96\relax
\mciteBstWouldAddEndPuncttrue
\mciteSetBstMidEndSepPunct{\mcitedefaultmidpunct}
{\mcitedefaultendpunct}{\mcitedefaultseppunct}\relax
\EndOfBibitem
\bibitem[Kohn and Sham(1965)Kohn, and Sham]{Kohn1965}
Kohn,~W.; Sham,~L.~J. Self-Consistent Equations Including Exchange and
  Correlation Effects. \emph{Phys. Rev.} \textbf{1965}, \emph{140},
  A1133--A1138\relax
\mciteBstWouldAddEndPuncttrue
\mciteSetBstMidEndSepPunct{\mcitedefaultmidpunct}
{\mcitedefaultendpunct}{\mcitedefaultseppunct}\relax
\EndOfBibitem
\bibitem[Mahan(1990)]{mahan-mpp}
Mahan,~G. \emph{Many-particle physics}; Plenum Press: New York, 1990\relax
\mciteBstWouldAddEndPuncttrue
\mciteSetBstMidEndSepPunct{\mcitedefaultmidpunct}
{\mcitedefaultendpunct}{\mcitedefaultseppunct}\relax
\EndOfBibitem
\bibitem[Luttinger and Ward(1960)Luttinger, and Ward]{Luttinger1960}
Luttinger,~J.~M.; Ward,~J.~C. Ground-State Energy of a Many-Fermion System.
  {II}. \emph{Phys. Rev.} \textbf{1960}, \emph{118}, 1417--1427\relax
\mciteBstWouldAddEndPuncttrue
\mciteSetBstMidEndSepPunct{\mcitedefaultmidpunct}
{\mcitedefaultendpunct}{\mcitedefaultseppunct}\relax
\EndOfBibitem
\bibitem[Klein(1961)]{Klein1961}
Klein,~A. Perturbation theory for an infinite medium of fermions. {II}.
  \emph{Phys. Rev.} \textbf{1961}, \emph{121}, 950--956\relax
\mciteBstWouldAddEndPuncttrue
\mciteSetBstMidEndSepPunct{\mcitedefaultmidpunct}
{\mcitedefaultendpunct}{\mcitedefaultseppunct}\relax
\EndOfBibitem
\bibitem[Holm(1999)]{Holm1999}
Holm,~B. Total Energies from $\mathit{GW}$ Calculations. \emph{Phys. Rev.
  Lett.} \textbf{1999}, \emph{83}, 788--791\relax
\mciteBstWouldAddEndPuncttrue
\mciteSetBstMidEndSepPunct{\mcitedefaultmidpunct}
{\mcitedefaultendpunct}{\mcitedefaultseppunct}\relax
\EndOfBibitem
\bibitem[Garc\'{\i}a-Gonz\'alez and Godby(2001)Garc\'{\i}a-Gonz\'alez, and
  Godby]{Garcia2001}
Garc\'{\i}a-Gonz\'alez,~P.; Godby,~R.~W. Self-consistent calculation of total
  energies of the electron gas using many-body perturbation theory. \emph{Phys.
  Rev. B} \textbf{2001}, \emph{63}, 075112\relax
\mciteBstWouldAddEndPuncttrue
\mciteSetBstMidEndSepPunct{\mcitedefaultmidpunct}
{\mcitedefaultendpunct}{\mcitedefaultseppunct}\relax
\EndOfBibitem
\bibitem[Caruso \latin{et~al.}(2012)Caruso, Rinke, Ren, Scheffler, and
  Rubio]{Caruso2012}
Caruso,~F.; Rinke,~P.; Ren,~X.; Scheffler,~M.; Rubio,~A. Unified description of
  ground and excited states of finite systems: The self-consistent $GW$
  approach. \emph{Phys. Rev. B} \textbf{2012}, \emph{86}, 081102\relax
\mciteBstWouldAddEndPuncttrue
\mciteSetBstMidEndSepPunct{\mcitedefaultmidpunct}
{\mcitedefaultendpunct}{\mcitedefaultseppunct}\relax
\EndOfBibitem
\bibitem[Bruneval \latin{et~al.}(2021)Bruneval, Rodriguez-Mayorga, Rinke, and
  Dvorak]{Bruneval2021b}
Bruneval,~F.; Rodriguez-Mayorga,~M.; Rinke,~P.; Dvorak,~M. Improved One-Shot
  Total Energies from the Linearized GW Density Matrix. \emph{Journal of
  Chemical Theory and Computation} \textbf{2021}, \emph{17}, 2126--2136, PMID:
  33705127\relax
\mciteBstWouldAddEndPuncttrue
\mciteSetBstMidEndSepPunct{\mcitedefaultmidpunct}
{\mcitedefaultendpunct}{\mcitedefaultseppunct}\relax
\EndOfBibitem
\bibitem[Garc\'{\i}a-Gonz\'alez and Godby(2002)Garc\'{\i}a-Gonz\'alez, and
  Godby]{Garcia2002}
Garc\'{\i}a-Gonz\'alez,~P.; Godby,~R.~W. Many-Body $\mathit{GW}$ Calculations
  of Ground-State Properties: Quasi-2D Electron Systems and van der Waals
  Forces. \emph{Phys. Rev. Lett.} \textbf{2002}, \emph{88}, 056406\relax
\mciteBstWouldAddEndPuncttrue
\mciteSetBstMidEndSepPunct{\mcitedefaultmidpunct}
{\mcitedefaultendpunct}{\mcitedefaultseppunct}\relax
\EndOfBibitem
\bibitem[Hedin(1965)]{Hedin1965}
Hedin,~L. New Method for Calculating the One-Particle Green's Function with
  Application to the Electron-Gas Problem. \emph{Phys. Rev.} \textbf{1965},
  \emph{139}, A796--A823\relax
\mciteBstWouldAddEndPuncttrue
\mciteSetBstMidEndSepPunct{\mcitedefaultmidpunct}
{\mcitedefaultendpunct}{\mcitedefaultseppunct}\relax
\EndOfBibitem
\bibitem[Hybertsen and Louie(1985)Hybertsen, and Louie]{Hybertsen1985}
Hybertsen,~M.~S.; Louie,~S.~G. First-Principles Theory of Quasiparticles:
  Calculation of Band Gaps in Semiconductors and Insulators. \emph{Phys. Rev.
  Lett.} \textbf{1985}, \emph{55}, 1418--1421\relax
\mciteBstWouldAddEndPuncttrue
\mciteSetBstMidEndSepPunct{\mcitedefaultmidpunct}
{\mcitedefaultendpunct}{\mcitedefaultseppunct}\relax
\EndOfBibitem
\bibitem[Godby \latin{et~al.}(1987)Godby, Schl\"uter, and Sham]{Godby1987}
Godby,~R.~W.; Schl\"uter,~M.; Sham,~L.~J. Trends in self-energy operators and
  their corresponding exchange-correlation potentials. \emph{Phys. Rev. B}
  \textbf{1987}, \emph{36}, 6497--6500\relax
\mciteBstWouldAddEndPuncttrue
\mciteSetBstMidEndSepPunct{\mcitedefaultmidpunct}
{\mcitedefaultendpunct}{\mcitedefaultseppunct}\relax
\EndOfBibitem
\bibitem[Godby \latin{et~al.}(1988)Godby, Schl\"uter, and Sham]{Godby1988}
Godby,~R.~W.; Schl\"uter,~M.; Sham,~L.~J. Self-energy operators and
  exchange-correlation potentials in semiconductors. \emph{Phys. Rev. B}
  \textbf{1988}, \emph{37}, 10159--10175\relax
\mciteBstWouldAddEndPuncttrue
\mciteSetBstMidEndSepPunct{\mcitedefaultmidpunct}
{\mcitedefaultendpunct}{\mcitedefaultseppunct}\relax
\EndOfBibitem
\bibitem[Blase \latin{et~al.}(1995)Blase, Rubio, Louie, and Cohen]{Blase1995}
Blase,~X.; Rubio,~A.; Louie,~S.~G.; Cohen,~M.~L. Quasiparticle band structure
  of bulk hexagonal boron nitride and related systems. \emph{Phys. Rev. B}
  \textbf{1995}, \emph{51}, 6868--6875\relax
\mciteBstWouldAddEndPuncttrue
\mciteSetBstMidEndSepPunct{\mcitedefaultmidpunct}
{\mcitedefaultendpunct}{\mcitedefaultseppunct}\relax
\EndOfBibitem
\bibitem[van Schilfgaarde \latin{et~al.}(2006)van Schilfgaarde, Kotani, and
  Faleev]{vSch2006}
van Schilfgaarde,~M.; Kotani,~T.; Faleev,~S.~V. Adequacy of approximations in
  $\mathit{GW}$ theory. \emph{Phys. Rev. B} \textbf{2006}, \emph{74},
  245125\relax
\mciteBstWouldAddEndPuncttrue
\mciteSetBstMidEndSepPunct{\mcitedefaultmidpunct}
{\mcitedefaultendpunct}{\mcitedefaultseppunct}\relax
\EndOfBibitem
\bibitem[Kotani \latin{et~al.}(2007)Kotani, van Schilfgaarde, and
  Faleev]{Kotani2007}
Kotani,~T.; van Schilfgaarde,~M.; Faleev,~S.~V. Quasiparticle self-consistent
  $GW$ method: A basis for the independent-particle approximation. \emph{Phys.
  Rev. B} \textbf{2007}, \emph{76}, 165106\relax
\mciteBstWouldAddEndPuncttrue
\mciteSetBstMidEndSepPunct{\mcitedefaultmidpunct}
{\mcitedefaultendpunct}{\mcitedefaultseppunct}\relax
\EndOfBibitem
\bibitem[Reining(2018)]{Reining2017}
Reining,~L. The GW approximation: content, successes and limitations.
  \emph{WIREs Computational Molecular Science} \textbf{2018}, \emph{8},
  e1344\relax
\mciteBstWouldAddEndPuncttrue
\mciteSetBstMidEndSepPunct{\mcitedefaultmidpunct}
{\mcitedefaultendpunct}{\mcitedefaultseppunct}\relax
\EndOfBibitem
\bibitem[Bruneval \latin{et~al.}(2021)Bruneval, Dattani, and van
  Setten]{Bruneval2021a}
Bruneval,~F.; Dattani,~N.; van Setten,~M.~J. The GW Miracle in Many-Body
  Perturbation Theory for the Ionization Potential of Molecules.
  \emph{Frontiers in Chemistry} \textbf{2021}, \emph{9}\relax
\mciteBstWouldAddEndPuncttrue
\mciteSetBstMidEndSepPunct{\mcitedefaultmidpunct}
{\mcitedefaultendpunct}{\mcitedefaultseppunct}\relax
\EndOfBibitem
\bibitem[Romaniello \latin{et~al.}(2009)Romaniello, Guyot, and
  Reining]{Romaniello2009}
Romaniello,~P.; Guyot,~S.; Reining,~L. The self-energy beyond GW: Local and
  nonlocal vertex corrections. \emph{The Journal of Chemical Physics}
  \textbf{2009}, \emph{131}, 154111\relax
\mciteBstWouldAddEndPuncttrue
\mciteSetBstMidEndSepPunct{\mcitedefaultmidpunct}
{\mcitedefaultendpunct}{\mcitedefaultseppunct}\relax
\EndOfBibitem
\bibitem[Aryasetiawan \latin{et~al.}(1996)Aryasetiawan, Hedin, and
  Karlsson]{Aryasetiawan1996}
Aryasetiawan,~F.; Hedin,~L.; Karlsson,~K. Multiple Plasmon Satellites in Na and
  Al Spectral Functions from Ab Initio Cumulant Expansion. \emph{Phys. Rev.
  Lett.} \textbf{1996}, \emph{77}, 2268--2271\relax
\mciteBstWouldAddEndPuncttrue
\mciteSetBstMidEndSepPunct{\mcitedefaultmidpunct}
{\mcitedefaultendpunct}{\mcitedefaultseppunct}\relax
\EndOfBibitem
\bibitem[Guzzo \latin{et~al.}(2011)Guzzo, Lani, Sottile, Romaniello, Gatti,
  Kas, Rehr, Silly, Sirotti, and Reining]{Guzzo2011}
Guzzo,~M.; Lani,~G.; Sottile,~F.; Romaniello,~P.; Gatti,~M.; Kas,~J.~J.;
  Rehr,~J.~J.; Silly,~M.~G.; Sirotti,~F.; Reining,~L. Valence Electron
  Photoemission Spectrum of Semiconductors: Ab Initio Description of Multiple
  Satellites. \emph{Phys. Rev. Lett.} \textbf{2011}, \emph{107}, 166401\relax
\mciteBstWouldAddEndPuncttrue
\mciteSetBstMidEndSepPunct{\mcitedefaultmidpunct}
{\mcitedefaultendpunct}{\mcitedefaultseppunct}\relax
\EndOfBibitem
\bibitem[Stan \latin{et~al.}(2009)Stan, Dahlen, and van Leeuwen]{Stan2009}
Stan,~A.; Dahlen,~N.~E.; van Leeuwen,~R. {Levels of self-consistency in the GW
  approximation}. \emph{The Journal of Chemical Physics} \textbf{2009},
  \emph{130}, 114105\relax
\mciteBstWouldAddEndPuncttrue
\mciteSetBstMidEndSepPunct{\mcitedefaultmidpunct}
{\mcitedefaultendpunct}{\mcitedefaultseppunct}\relax
\EndOfBibitem
\bibitem[Minnhagen(1975)]{Minnhagen1975}
Minnhagen,~P. Aspects on diagrammatic expansion for models related to a
  homogeneous electron gas. \emph{J. Phys. C: Solid State Phys.} \textbf{1975},
  \emph{8}, 1535\relax
\mciteBstWouldAddEndPuncttrue
\mciteSetBstMidEndSepPunct{\mcitedefaultmidpunct}
{\mcitedefaultendpunct}{\mcitedefaultseppunct}\relax
\EndOfBibitem
\bibitem[Bobbert and van Haeringen(1994)Bobbert, and van
  Haeringen]{Bobbert1994}
Bobbert,~P.~A.; van Haeringen,~W. Lowest-order vertex-correction contribution
  to the direct gap of silicon. \emph{Phys. Rev. B} \textbf{1994}, \emph{49},
  10326--10331\relax
\mciteBstWouldAddEndPuncttrue
\mciteSetBstMidEndSepPunct{\mcitedefaultmidpunct}
{\mcitedefaultendpunct}{\mcitedefaultseppunct}\relax
\EndOfBibitem
\bibitem[Shirley(1996)]{Shirley1996}
Shirley,~E.~L. Self-consistent GW and higher-order calculations of electron
  states in metals. \emph{Phys. Rev. B} \textbf{1996}, \emph{54},
  7758--7764\relax
\mciteBstWouldAddEndPuncttrue
\mciteSetBstMidEndSepPunct{\mcitedefaultmidpunct}
{\mcitedefaultendpunct}{\mcitedefaultseppunct}\relax
\EndOfBibitem
\bibitem[Gr\"uneis \latin{et~al.}(2014)Gr\"uneis, Kresse, Hinuma, and
  Oba]{Gruneis2014}
Gr\"uneis,~A.; Kresse,~G.; Hinuma,~Y.; Oba,~F. Ionization Potentials of Solids:
  The Importance of Vertex Corrections. \emph{Phys. Rev. Lett.} \textbf{2014},
  \emph{112}, 096401\relax
\mciteBstWouldAddEndPuncttrue
\mciteSetBstMidEndSepPunct{\mcitedefaultmidpunct}
{\mcitedefaultendpunct}{\mcitedefaultseppunct}\relax
\EndOfBibitem
\bibitem[Hinuma \latin{et~al.}(2014)Hinuma, Gr\"uneis, Kresse, and
  Oba]{Hinuma2014}
Hinuma,~Y.; Gr\"uneis,~A.; Kresse,~G.; Oba,~F. Band alignment of semiconductors
  from density-functional theory and many-body perturbation theory. \emph{Phys.
  Rev. B} \textbf{2014}, \emph{90}, 155405\relax
\mciteBstWouldAddEndPuncttrue
\mciteSetBstMidEndSepPunct{\mcitedefaultmidpunct}
{\mcitedefaultendpunct}{\mcitedefaultseppunct}\relax
\EndOfBibitem
\bibitem[Ren \latin{et~al.}(2015)Ren, Marom, Caruso, Scheffler, and
  Rinke]{Ren2015}
Ren,~X.; Marom,~N.; Caruso,~F.; Scheffler,~M.; Rinke,~P. Beyond the $GW$
  approximation: A second-order screened exchange correction. \emph{Phys. Rev.
  B} \textbf{2015}, \emph{92}, 081104\relax
\mciteBstWouldAddEndPuncttrue
\mciteSetBstMidEndSepPunct{\mcitedefaultmidpunct}
{\mcitedefaultendpunct}{\mcitedefaultseppunct}\relax
\EndOfBibitem
\bibitem[Kutepov(2016)]{Kutepov2016}
Kutepov,~A.~L. Electronic structure of Na, K, Si, and LiF from self-consistent
  solution of Hedin's equations including vertex corrections. \emph{Phys. Rev.
  B} \textbf{2016}, \emph{94}, 155101\relax
\mciteBstWouldAddEndPuncttrue
\mciteSetBstMidEndSepPunct{\mcitedefaultmidpunct}
{\mcitedefaultendpunct}{\mcitedefaultseppunct}\relax
\EndOfBibitem
\bibitem[Kutepov(2017)]{Kutepov2017}
Kutepov,~A.~L. Self-consistent solution of Hedin's equations: Semiconductors
  and insulators. \emph{Phys. Rev. B} \textbf{2017}, \emph{95}, 195120\relax
\mciteBstWouldAddEndPuncttrue
\mciteSetBstMidEndSepPunct{\mcitedefaultmidpunct}
{\mcitedefaultendpunct}{\mcitedefaultseppunct}\relax
\EndOfBibitem
\bibitem[Pavlyukh \latin{et~al.}(2016)Pavlyukh, Uimonen, Stefanucci, and van
  Leeuwen]{Pavlyukh2016}
Pavlyukh,~Y.; Uimonen,~A.-M.; Stefanucci,~G.; van Leeuwen,~R. Vertex
  Corrections for Positive-Definite Spectral Functions of Simple Metals.
  \emph{Phys. Rev. Lett.} \textbf{2016}, \emph{117}, 206402\relax
\mciteBstWouldAddEndPuncttrue
\mciteSetBstMidEndSepPunct{\mcitedefaultmidpunct}
{\mcitedefaultendpunct}{\mcitedefaultseppunct}\relax
\EndOfBibitem
\bibitem[Maggio and Kresse(2017)Maggio, and Kresse]{Maggio2017}
Maggio,~E.; Kresse,~G. GW Vertex Corrected Calculations for Molecular Systems.
  \emph{Journal of Chemical Theory and Computation} \textbf{2017}, \emph{13},
  4765--4778, PMID: 28873298\relax
\mciteBstWouldAddEndPuncttrue
\mciteSetBstMidEndSepPunct{\mcitedefaultmidpunct}
{\mcitedefaultendpunct}{\mcitedefaultseppunct}\relax
\EndOfBibitem
\bibitem[Wang and Ren(2022)Wang, and Ren]{Wang2022}
Wang,~Y.; Ren,~X. {Vertex effects in describing the ionization energies of the
  first-row transition-metal monoxide molecules}. \emph{The Journal of Chemical
  Physics} \textbf{2022}, \emph{157}, 214115\relax
\mciteBstWouldAddEndPuncttrue
\mciteSetBstMidEndSepPunct{\mcitedefaultmidpunct}
{\mcitedefaultendpunct}{\mcitedefaultseppunct}\relax
\EndOfBibitem
\bibitem[Langreth and Perdew(1975)Langreth, and Perdew]{Langreth1975}
Langreth,~D.; Perdew,~J. The exchange-correlation energy of a metallic surface.
  \emph{Solid State Communications} \textbf{1975}, \emph{17}, 1425--1429\relax
\mciteBstWouldAddEndPuncttrue
\mciteSetBstMidEndSepPunct{\mcitedefaultmidpunct}
{\mcitedefaultendpunct}{\mcitedefaultseppunct}\relax
\EndOfBibitem
\bibitem[Langreth and Perdew(1977)Langreth, and Perdew]{Langreth1977}
Langreth,~D.~C.; Perdew,~J.~P. Exchange-correlation energy of a metallic
  surface: Wave-vector analysis. \emph{Phys. Rev. B} \textbf{1977}, \emph{15},
  2884--2901\relax
\mciteBstWouldAddEndPuncttrue
\mciteSetBstMidEndSepPunct{\mcitedefaultmidpunct}
{\mcitedefaultendpunct}{\mcitedefaultseppunct}\relax
\EndOfBibitem
\bibitem[Runge and Gross(1984)Runge, and Gross]{Gross1984}
Runge,~E.; Gross,~E. K.~U. Density-Functional Theory for Time-Dependent
  Systems. \emph{Phys. Rev. Lett.} \textbf{1984}, \emph{52}, 997--1000\relax
\mciteBstWouldAddEndPuncttrue
\mciteSetBstMidEndSepPunct{\mcitedefaultmidpunct}
{\mcitedefaultendpunct}{\mcitedefaultseppunct}\relax
\EndOfBibitem
\bibitem[Del~Sole \latin{et~al.}(1994)Del~Sole, Reining, and
  Godby]{DelSole1994}
Del~Sole,~R.; Reining,~L.; Godby,~R.~W. GW\ensuremath{\Gamma} approximation for
  electron self-energies in semiconductors and insulators. \emph{Phys. Rev. B}
  \textbf{1994}, \emph{49}, 8024--8028\relax
\mciteBstWouldAddEndPuncttrue
\mciteSetBstMidEndSepPunct{\mcitedefaultmidpunct}
{\mcitedefaultendpunct}{\mcitedefaultseppunct}\relax
\EndOfBibitem
\bibitem[Reining \latin{et~al.}(2002)Reining, Olevano, Rubio, and
  Onida]{Reining2002}
Reining,~L.; Olevano,~V.; Rubio,~A.; Onida,~G. Excitonic Effects in Solids
  Described by Time-Dependent Density-Functional Theory. \emph{Phys. Rev.
  Lett.} \textbf{2002}, \emph{88}, 066404\relax
\mciteBstWouldAddEndPuncttrue
\mciteSetBstMidEndSepPunct{\mcitedefaultmidpunct}
{\mcitedefaultendpunct}{\mcitedefaultseppunct}\relax
\EndOfBibitem
\bibitem[Overhauser(1971)]{Overhauser1971}
Overhauser,~A.~W. Simplified Theory of Electron Correlations in Metals.
  \emph{Phys. Rev. B} \textbf{1971}, \emph{3}, 1888--1898\relax
\mciteBstWouldAddEndPuncttrue
\mciteSetBstMidEndSepPunct{\mcitedefaultmidpunct}
{\mcitedefaultendpunct}{\mcitedefaultseppunct}\relax
\EndOfBibitem
\bibitem[Petrillo and Sacchetti(1988)Petrillo, and Sacchetti]{Petrillo1988}
Petrillo,~C.; Sacchetti,~F. Electron-gas self-energy at metallic density.
  \emph{Phys. Rev. B} \textbf{1988}, \emph{38}, 3834--3840\relax
\mciteBstWouldAddEndPuncttrue
\mciteSetBstMidEndSepPunct{\mcitedefaultmidpunct}
{\mcitedefaultendpunct}{\mcitedefaultseppunct}\relax
\EndOfBibitem
\bibitem[Mahan and Sernelius(1989)Mahan, and Sernelius]{Mahan1989}
Mahan,~G.~D.; Sernelius,~B.~E. Electron-electron interactions and the bandwidth
  of metals. \emph{Phys. Rev. Lett.} \textbf{1989}, \emph{62}, 2718--2720\relax
\mciteBstWouldAddEndPuncttrue
\mciteSetBstMidEndSepPunct{\mcitedefaultmidpunct}
{\mcitedefaultendpunct}{\mcitedefaultseppunct}\relax
\EndOfBibitem
\bibitem[Hybertsen and Louie(1986)Hybertsen, and Louie]{Hybertsen1986}
Hybertsen,~M.~S.; Louie,~S.~G. Electron correlation in semiconductors and
  insulators: Band gaps and quasiparticle energies. \emph{Phys. Rev. B}
  \textbf{1986}, \emph{34}, 5390--5413\relax
\mciteBstWouldAddEndPuncttrue
\mciteSetBstMidEndSepPunct{\mcitedefaultmidpunct}
{\mcitedefaultendpunct}{\mcitedefaultseppunct}\relax
\EndOfBibitem
\bibitem[Hindgren and Almbladh(1997)Hindgren, and Almbladh]{Hindgren1997}
Hindgren,~M.; Almbladh,~C.-O. Improved local-field corrections to the
  ${G}_{0}W$ approximation in jellium: Importance of consistency relations.
  \emph{Phys. Rev. B} \textbf{1997}, \emph{56}, 12832--12839\relax
\mciteBstWouldAddEndPuncttrue
\mciteSetBstMidEndSepPunct{\mcitedefaultmidpunct}
{\mcitedefaultendpunct}{\mcitedefaultseppunct}\relax
\EndOfBibitem
\bibitem[Schmidt \latin{et~al.}(2017)Schmidt, Patrick, and
  Thygesen]{Schmidt2017}
Schmidt,~P.~S.; Patrick,~C.~E.; Thygesen,~K.~S. Simple vertex correction
  improves $GW$ band energies of bulk and two-dimensional crystals. \emph{Phys.
  Rev. B} \textbf{2017}, \emph{96}, 205206\relax
\mciteBstWouldAddEndPuncttrue
\mciteSetBstMidEndSepPunct{\mcitedefaultmidpunct}
{\mcitedefaultendpunct}{\mcitedefaultseppunct}\relax
\EndOfBibitem
\bibitem[Hung \latin{et~al.}(2016)Hung, da~Jornada, Souto-Casares, Chelikowsky,
  Louie, and \"O\ifmmode~\breve{g}\else \u{g}\fi{}\"ut]{Hung2016}
Hung,~L.; da~Jornada,~F.~H.; Souto-Casares,~J.; Chelikowsky,~J.~R.;
  Louie,~S.~G.; \"O\ifmmode~\breve{g}\else \u{g}\fi{}\"ut,~S. Excitation
  spectra of aromatic molecules within a real-space $GW$-BSE formalism: Role of
  self-consistency and vertex corrections. \emph{Phys. Rev. B} \textbf{2016},
  \emph{94}, 085125\relax
\mciteBstWouldAddEndPuncttrue
\mciteSetBstMidEndSepPunct{\mcitedefaultmidpunct}
{\mcitedefaultendpunct}{\mcitedefaultseppunct}\relax
\EndOfBibitem
\bibitem[Olsen \latin{et~al.}(2019)Olsen, Patrick, Bates, Ruzsinszky, and
  Thygesen]{Olsen2019}
Olsen,~T.; Patrick,~C.~E.; Bates,~J.~E.; Ruzsinszky,~A.; Thygesen,~K.~S. Beyond
  the RPA and GW methods with adiabatic xc-kernels for accurate ground state
  and quasiparticle energies. \emph{npj Computational Materials} \textbf{2019},
  \emph{5}, 106\relax
\mciteBstWouldAddEndPuncttrue
\mciteSetBstMidEndSepPunct{\mcitedefaultmidpunct}
{\mcitedefaultendpunct}{\mcitedefaultseppunct}\relax
\EndOfBibitem
\bibitem[Chen and Pasquarello(2015)Chen, and Pasquarello]{Chen2015}
Chen,~W.; Pasquarello,~A. Accurate band gaps of extended systems via efficient
  vertex corrections in $GW$. \emph{Phys. Rev. B} \textbf{2015}, \emph{92},
  041115\relax
\mciteBstWouldAddEndPuncttrue
\mciteSetBstMidEndSepPunct{\mcitedefaultmidpunct}
{\mcitedefaultendpunct}{\mcitedefaultseppunct}\relax
\EndOfBibitem
\bibitem[Tal \latin{et~al.}(2021)Tal, Chen, and Pasquarello]{Tal2021}
Tal,~A.; Chen,~W.; Pasquarello,~A. Vertex function compliant with the Ward
  identity for quasiparticle self-consistent calculations beyond $GW$.
  \emph{Phys. Rev. B} \textbf{2021}, \emph{103}, L161104\relax
\mciteBstWouldAddEndPuncttrue
\mciteSetBstMidEndSepPunct{\mcitedefaultmidpunct}
{\mcitedefaultendpunct}{\mcitedefaultseppunct}\relax
\EndOfBibitem
\bibitem[Gross and Kohn(1985)Gross, and Kohn]{Gross1985}
Gross,~E. K.~U.; Kohn,~W. Local density-functional theory of
  frequency-dependent linear response. \emph{Phys. Rev. Lett.} \textbf{1985},
  \emph{55}, 2850--2852\relax
\mciteBstWouldAddEndPuncttrue
\mciteSetBstMidEndSepPunct{\mcitedefaultmidpunct}
{\mcitedefaultendpunct}{\mcitedefaultseppunct}\relax
\EndOfBibitem
\bibitem[Tokatly and Pankratov(2001)Tokatly, and Pankratov]{Tokatly2001}
Tokatly,~I.~V.; Pankratov,~O. Many-Body Diagrammatic Expansion in a Kohn-Sham
  Basis: Implications for Time-Dependent Density Functional Theory of Excited
  States. \emph{Phys. Rev. Lett.} \textbf{2001}, \emph{86}, 2078--2081\relax
\mciteBstWouldAddEndPuncttrue
\mciteSetBstMidEndSepPunct{\mcitedefaultmidpunct}
{\mcitedefaultendpunct}{\mcitedefaultseppunct}\relax
\EndOfBibitem
\bibitem[Bruneval \latin{et~al.}(2005)Bruneval, Sottile, Olevano, Del~Sole, and
  Reining]{Bruneval2005}
Bruneval,~F.; Sottile,~F.; Olevano,~V.; Del~Sole,~R.; Reining,~L. Many-Body
  Perturbation Theory Using the Density-Functional Concept: Beyond the $GW$
  Approximation. \emph{Phys. Rev. Lett.} \textbf{2005}, \emph{94}, 186402\relax
\mciteBstWouldAddEndPuncttrue
\mciteSetBstMidEndSepPunct{\mcitedefaultmidpunct}
{\mcitedefaultendpunct}{\mcitedefaultseppunct}\relax
\EndOfBibitem
\bibitem[Gatti \latin{et~al.}(2007)Gatti, Olevano, Reining, and
  Tokatly]{Gatti2007}
Gatti,~M.; Olevano,~V.; Reining,~L.; Tokatly,~I.~V. Transforming Nonlocality
  into a Frequency Dependence: A Shortcut to Spectroscopy. \emph{Phys. Rev.
  Lett.} \textbf{2007}, \emph{99}, 057401\relax
\mciteBstWouldAddEndPuncttrue
\mciteSetBstMidEndSepPunct{\mcitedefaultmidpunct}
{\mcitedefaultendpunct}{\mcitedefaultseppunct}\relax
\EndOfBibitem
\bibitem[Botti \latin{et~al.}(2007)Botti, Schindlmayr, Sole, and
  Reining]{Botti2007}
Botti,~S.; Schindlmayr,~A.; Sole,~R.~D.; Reining,~L. Time-dependent
  density-functional theory for extended systems. \emph{Reports on Progress in
  Physics} \textbf{2007}, \emph{70}, 357\relax
\mciteBstWouldAddEndPuncttrue
\mciteSetBstMidEndSepPunct{\mcitedefaultmidpunct}
{\mcitedefaultendpunct}{\mcitedefaultseppunct}\relax
\EndOfBibitem
\bibitem[Singhal and Callaway(1976)Singhal, and Callaway]{Singhal1976}
Singhal,~S.~P.; Callaway,~J. Exchange correction to the dielectric function in
  the local exchange approximation. \emph{Phys. Rev. B} \textbf{1976},
  \emph{14}, 2347--2351\relax
\mciteBstWouldAddEndPuncttrue
\mciteSetBstMidEndSepPunct{\mcitedefaultmidpunct}
{\mcitedefaultendpunct}{\mcitedefaultseppunct}\relax
\EndOfBibitem
\bibitem[Hybertsen and Louie(1987)Hybertsen, and Louie]{Hybertsen1987}
Hybertsen,~M.~S.; Louie,~S.~G. Ab initio static dielectric matrices from the
  density-functional approach. I. Formulation and application to semiconductors
  and insulators. \emph{Phys. Rev. B} \textbf{1987}, \emph{35},
  5585--5601\relax
\mciteBstWouldAddEndPuncttrue
\mciteSetBstMidEndSepPunct{\mcitedefaultmidpunct}
{\mcitedefaultendpunct}{\mcitedefaultseppunct}\relax
\EndOfBibitem
\bibitem[Hellgren and Baguet(2023)Hellgren, and Baguet]{Hellgren2023}
Hellgren,~M.; Baguet,~L. {Strengths and limitations of the adiabatic
  exact-exchange kernel for total energy calculations}. \emph{The Journal of
  Chemical Physics} \textbf{2023}, \emph{158}, 184107\relax
\mciteBstWouldAddEndPuncttrue
\mciteSetBstMidEndSepPunct{\mcitedefaultmidpunct}
{\mcitedefaultendpunct}{\mcitedefaultseppunct}\relax
\EndOfBibitem
\bibitem[Sottile \latin{et~al.}(2003)Sottile, Olevano, and
  Reining]{Sottile2003}
Sottile,~F.; Olevano,~V.; Reining,~L. \emph{Phys. Rev. Lett.} \textbf{2003},
  \emph{91}, 056402\relax
\mciteBstWouldAddEndPuncttrue
\mciteSetBstMidEndSepPunct{\mcitedefaultmidpunct}
{\mcitedefaultendpunct}{\mcitedefaultseppunct}\relax
\EndOfBibitem
\bibitem[Sharma \latin{et~al.}(2011)Sharma, Dewhurst, Sanna, and
  Gross]{Sharma2011}
Sharma,~S.; Dewhurst,~J.~K.; Sanna,~A.; Gross,~E. K.~U. Bootstrap Approximation
  for the Exchange-Correlation Kernel of Time-Dependent Density-Functional
  Theory. \emph{Phys. Rev. Lett.} \textbf{2011}, \emph{107}\relax
\mciteBstWouldAddEndPuncttrue
\mciteSetBstMidEndSepPunct{\mcitedefaultmidpunct}
{\mcitedefaultendpunct}{\mcitedefaultseppunct}\relax
\EndOfBibitem
\bibitem[Adragna \latin{et~al.}(2003)Adragna, {Del Sole}, and
  Marini]{Adragna2003}
Adragna,~G.; {Del Sole},~R.; Marini,~A. \emph{Phys. Rev. B} \textbf{2003},
  \emph{68}, 165108\relax
\mciteBstWouldAddEndPuncttrue
\mciteSetBstMidEndSepPunct{\mcitedefaultmidpunct}
{\mcitedefaultendpunct}{\mcitedefaultseppunct}\relax
\EndOfBibitem
\bibitem[Marini \latin{et~al.}(2003)Marini, {Del Sole}, and Rubio]{Marini2003}
Marini,~A.; {Del Sole},~R.; Rubio,~A. \emph{Phys. Rev. Lett.} \textbf{2003},
  \emph{91}, 256402\relax
\mciteBstWouldAddEndPuncttrue
\mciteSetBstMidEndSepPunct{\mcitedefaultmidpunct}
{\mcitedefaultendpunct}{\mcitedefaultseppunct}\relax
\EndOfBibitem
\bibitem[Rigamonti \latin{et~al.}(2015)Rigamonti, Botti, Veniard, Draxl,
  Reining, and Sottile]{Rigamonti2015}
Rigamonti,~S.; Botti,~S.; Veniard,~V.; Draxl,~C.; Reining,~L.; Sottile,~F.
  Estimating Excitonic Effects in the Absorption Spectra of Solids: Problems
  and Insight from a Guided Iteration Scheme. \emph{Phys. Rev. Lett.}
  \textbf{2015}, \emph{114}, 146402\relax
\mciteBstWouldAddEndPuncttrue
\mciteSetBstMidEndSepPunct{\mcitedefaultmidpunct}
{\mcitedefaultendpunct}{\mcitedefaultseppunct}\relax
\EndOfBibitem
\bibitem[Levy and Perdew(1985)Levy, and Perdew]{Levy1985}
Levy,~M.; Perdew,~J.~P. Hellmann-Feynman, virial, and scaling requisites for
  the exact universal density functionals. Shape of the correlation potential
  and diamagnetic susceptibility for atoms. \emph{Phys. Rev. A} \textbf{1985},
  \emph{32}, 2010--2021\relax
\mciteBstWouldAddEndPuncttrue
\mciteSetBstMidEndSepPunct{\mcitedefaultmidpunct}
{\mcitedefaultendpunct}{\mcitedefaultseppunct}\relax
\EndOfBibitem
\bibitem[Jiang \latin{et~al.}(2020)Jiang, Mosquera, Oueis, and
  Wasserman]{Jiang2020}
Jiang,~K.; Mosquera,~M.~A.; Oueis,~Y.; Wasserman,~A. Virial relations in
  density embedding. \emph{International Journal of Quantum Chemistry}
  \textbf{2020}, \emph{120}, e26204\relax
\mciteBstWouldAddEndPuncttrue
\mciteSetBstMidEndSepPunct{\mcitedefaultmidpunct}
{\mcitedefaultendpunct}{\mcitedefaultseppunct}\relax
\EndOfBibitem
\bibitem[Kim \latin{et~al.}(2013)Kim, Sim, and Burke]{Kim2013}
Kim,~M.-C.; Sim,~E.; Burke,~K. Understanding and Reducing Errors in Density
  Functional Calculations. \emph{Phys. Rev. Lett.} \textbf{2013}, \emph{111},
  073003\relax
\mciteBstWouldAddEndPuncttrue
\mciteSetBstMidEndSepPunct{\mcitedefaultmidpunct}
{\mcitedefaultendpunct}{\mcitedefaultseppunct}\relax
\EndOfBibitem
\bibitem[Ren \latin{et~al.}(2012)Ren, Rinke, Joas, and Scheffler]{Ren2012}
Ren,~X.; Rinke,~P.; Joas,~C.; Scheffler,~M. Random-phase approximation and its
  applications in computational chemistry and materials science. \emph{Journal
  of Materials Science} \textbf{2012}, \emph{47}, 7447\relax
\mciteBstWouldAddEndPuncttrue
\mciteSetBstMidEndSepPunct{\mcitedefaultmidpunct}
{\mcitedefaultendpunct}{\mcitedefaultseppunct}\relax
\EndOfBibitem
\bibitem[Savin \latin{et~al.}(2001)Savin, Colonna, and Allavena]{Savin2001}
Savin,~A.; Colonna,~F.; Allavena,~M. {Analysis of the linear response function
  along the adiabatic connection from the Kohn–Sham to the correlated
  system}. \emph{The Journal of Chemical Physics} \textbf{2001}, \emph{115},
  6827--6833\relax
\mciteBstWouldAddEndPuncttrue
\mciteSetBstMidEndSepPunct{\mcitedefaultmidpunct}
{\mcitedefaultendpunct}{\mcitedefaultseppunct}\relax
\EndOfBibitem
\bibitem[Romaniello \latin{et~al.}(2012)Romaniello, Bechstedt, and
  Reining]{Romaniello2012}
Romaniello,~P.; Bechstedt,~F.; Reining,~L. Beyond the $GW$ approximation:
  Combining correlation channels. \emph{Phys. Rev. B} \textbf{2012}, \emph{85},
  155131\relax
\mciteBstWouldAddEndPuncttrue
\mciteSetBstMidEndSepPunct{\mcitedefaultmidpunct}
{\mcitedefaultendpunct}{\mcitedefaultseppunct}\relax
\EndOfBibitem
\bibitem[Carrascal \latin{et~al.}(2015)Carrascal, Ferrer, Smith, and
  Burke]{Carrascal2015}
Carrascal,~D.~J.; Ferrer,~J.; Smith,~J.~C.; Burke,~K. The Hubbard dimer: a
  density functional case study of a many-body problem. \emph{Journal of
  Physics: Condensed Matter} \textbf{2015}, \emph{27}, 393001\relax
\mciteBstWouldAddEndPuncttrue
\mciteSetBstMidEndSepPunct{\mcitedefaultmidpunct}
{\mcitedefaultendpunct}{\mcitedefaultseppunct}\relax
\EndOfBibitem
\bibitem[Aryasetiawan and Gunnarsson(2002)Aryasetiawan, and
  Gunnarsson]{Aryasetiawan2002}
Aryasetiawan,~F.; Gunnarsson,~O. Exchange-correlation kernel in time-dependent
  density functional theory. \emph{Phys. Rev. B} \textbf{2002}, \emph{66},
  165119\relax
\mciteBstWouldAddEndPuncttrue
\mciteSetBstMidEndSepPunct{\mcitedefaultmidpunct}
{\mcitedefaultendpunct}{\mcitedefaultseppunct}\relax
\EndOfBibitem
\bibitem[Coveney and Tew(2023)Coveney, and Tew]{Coveney2023}
Coveney,~C. J.~N.; Tew,~D.~P. A Regularized Second-Order Correlation Method
  from Green’s Function Theory. \emph{Journal of Chemical Theory and
  Computation} \textbf{2023}, \emph{19}, 3915–3928\relax
\mciteBstWouldAddEndPuncttrue
\mciteSetBstMidEndSepPunct{\mcitedefaultmidpunct}
{\mcitedefaultendpunct}{\mcitedefaultseppunct}\relax
\EndOfBibitem
\bibitem[J.Hubbard(1963)]{Hubbard1963}
J.Hubbard, {Electron correlations in narrow energy bands}. \emph{{Proc. R. Soc.
  Lond. A}} \textbf{1963}, \emph{276}, 238\relax
\mciteBstWouldAddEndPuncttrue
\mciteSetBstMidEndSepPunct{\mcitedefaultmidpunct}
{\mcitedefaultendpunct}{\mcitedefaultseppunct}\relax
\EndOfBibitem
\bibitem[J.Hubbard(1964)]{Hubbard1964}
J.Hubbard, {Electron correlations in narrow energy bands}. \emph{{Proc. R. Soc.
  Lond. A}} \textbf{1964}, \emph{277}, 237\relax
\mciteBstWouldAddEndPuncttrue
\mciteSetBstMidEndSepPunct{\mcitedefaultmidpunct}
{\mcitedefaultendpunct}{\mcitedefaultseppunct}\relax
\EndOfBibitem
\bibitem[Spataru \latin{et~al.}(2004)Spataru, Benedict, and Louie]{Spataru2004}
Spataru,~C.~D.; Benedict,~L.~X.; Louie,~S.~G. Ab initio calculation of band-gap
  renormalization in highly excited GaAs. \emph{Phys. Rev. B} \textbf{2004},
  \emph{69}, 205204\relax
\mciteBstWouldAddEndPuncttrue
\mciteSetBstMidEndSepPunct{\mcitedefaultmidpunct}
{\mcitedefaultendpunct}{\mcitedefaultseppunct}\relax
\EndOfBibitem
\bibitem[Honet \latin{et~al.}(2022)Honet, Henrard, and Meunier]{Honet2022}
Honet,~A.; Henrard,~L.; Meunier,~V. {Exact and many-body perturbation solutions
  of the Hubbard model applied to linear chains}. \emph{AIP Advances}
  \textbf{2022}, \emph{12}, 035238\relax
\mciteBstWouldAddEndPuncttrue
\mciteSetBstMidEndSepPunct{\mcitedefaultmidpunct}
{\mcitedefaultendpunct}{\mcitedefaultseppunct}\relax
\EndOfBibitem
\bibitem[Holm and von Barth(1998)Holm, and von Barth]{Holm1998}
Holm,~B.; von Barth,~U. Fully self-consistent $\mathrm{GW}$ self-energy of the
  electron gas. \emph{Phys. Rev. B} \textbf{1998}, \emph{57}, 2108--2117\relax
\mciteBstWouldAddEndPuncttrue
\mciteSetBstMidEndSepPunct{\mcitedefaultmidpunct}
{\mcitedefaultendpunct}{\mcitedefaultseppunct}\relax
\EndOfBibitem
\bibitem[Perdew \latin{et~al.}(2017)Perdew, Yang, Burke, Yang, Gross,
  Scheffler, Scuseria, Henderson, Zhang, Ruzsinszky, Peng, Sun, Trushin, and
  Görling]{Perdew2017}
Perdew,~J.~P.; Yang,~W.; Burke,~K.; Yang,~Z.; Gross,~E. K.~U.; Scheffler,~M.;
  Scuseria,~G.~E.; Henderson,~T.~M.; Zhang,~I.~Y.; Ruzsinszky,~A.; Peng,~H.;
  Sun,~J.; Trushin,~E.; Görling,~A. Understanding band gaps of solids in
  generalized Kohn–Sham theory. \emph{Proceedings of the National Academy of
  Sciences} \textbf{2017}, \emph{114}, 2801--2806\relax
\mciteBstWouldAddEndPuncttrue
\mciteSetBstMidEndSepPunct{\mcitedefaultmidpunct}
{\mcitedefaultendpunct}{\mcitedefaultseppunct}\relax
\EndOfBibitem
\bibitem[Heyd \latin{et~al.}(2003)Heyd, Scuseria, and Ernzerhof]{Heyd2003}
Heyd,~J.; Scuseria,~G.~E.; Ernzerhof,~M. Hybrid functionals based on a screened
  Coulomb potential. \emph{The Journal of Chemical Physics} \textbf{2003},
  \emph{118}, 8207--8215\relax
\mciteBstWouldAddEndPuncttrue
\mciteSetBstMidEndSepPunct{\mcitedefaultmidpunct}
{\mcitedefaultendpunct}{\mcitedefaultseppunct}\relax
\EndOfBibitem
\bibitem[Aryasetiawan \latin{et~al.}(2012)Aryasetiawan, Sakuma, and
  Karlsson]{Ferdi2012}
Aryasetiawan,~F.; Sakuma,~R.; Karlsson,~K. $GW$ approximation with
  self-screening correction. \emph{Phys. Rev. B} \textbf{2012}, \emph{85},
  035106\relax
\mciteBstWouldAddEndPuncttrue
\mciteSetBstMidEndSepPunct{\mcitedefaultmidpunct}
{\mcitedefaultendpunct}{\mcitedefaultseppunct}\relax
\EndOfBibitem
\bibitem[Hedin \latin{et~al.}(1967)Hedin, Lundqvist, and Lundqvist]{Hedin1967b}
Hedin,~L.; Lundqvist,~B.; Lundqvist,~S. New structure in the single-particle
  spectrum of an electron gas. \emph{Solid State Communications} \textbf{1967},
  \emph{5}, 237 -- 239\relax
\mciteBstWouldAddEndPuncttrue
\mciteSetBstMidEndSepPunct{\mcitedefaultmidpunct}
{\mcitedefaultendpunct}{\mcitedefaultseppunct}\relax
\EndOfBibitem
\bibitem[Bergerse.B \latin{et~al.}(1973)Bergerse.B, Kus, and
  Blomberg]{Bergerse1973}
Bergerse.B,; Kus,~F.~W.; Blomberg,~C. SINGLE-PARTICLE {G}REENS FUNCTION IN
  ELECTRON-PLASMON APPROXIMATION. \emph{Canadian J. Phys.} \textbf{1973},
  \emph{51}, 102--110\relax
\mciteBstWouldAddEndPuncttrue
\mciteSetBstMidEndSepPunct{\mcitedefaultmidpunct}
{\mcitedefaultendpunct}{\mcitedefaultseppunct}\relax
\EndOfBibitem
\bibitem[Guzzo \latin{et~al.}(2014)Guzzo, Kas, Sponza, Giorgetti, Sottile,
  Pierucci, Silly, Sirotti, Rehr, and Reining]{Guzzo2014}
Guzzo,~M.; Kas,~J.~J.; Sponza,~L.; Giorgetti,~C.; Sottile,~F.; Pierucci,~D.;
  Silly,~M.~G.; Sirotti,~F.; Rehr,~J.~J.; Reining,~L. Multiple satellites in
  materials with complex plasmon spectra: From graphite to graphene.
  \emph{Phys. Rev. B} \textbf{2014}, \emph{89}, 085425\relax
\mciteBstWouldAddEndPuncttrue
\mciteSetBstMidEndSepPunct{\mcitedefaultmidpunct}
{\mcitedefaultendpunct}{\mcitedefaultseppunct}\relax
\EndOfBibitem
\bibitem[Nelson \latin{et~al.}(2007)Nelson, Bokes, Rinke, and
  Godby]{Nelson2007}
Nelson,~W.; Bokes,~P.; Rinke,~P.; Godby,~R.~W. Self-interaction in
  Green's-function theory of the hydrogen atom. \emph{Phys. Rev. A}
  \textbf{2007}, \emph{75}, 032505\relax
\mciteBstWouldAddEndPuncttrue
\mciteSetBstMidEndSepPunct{\mcitedefaultmidpunct}
{\mcitedefaultendpunct}{\mcitedefaultseppunct}\relax
\EndOfBibitem
\bibitem[Fernandez(2009)]{Fernandez2009}
Fernandez,~J.~J. $GW$ calculations in an exactly solvable model system at
  different dilution regimes: The problem of the self-interaction in the
  correlation part. \emph{Phys. Rev. A} \textbf{2009}, \emph{79}, 052513\relax
\mciteBstWouldAddEndPuncttrue
\mciteSetBstMidEndSepPunct{\mcitedefaultmidpunct}
{\mcitedefaultendpunct}{\mcitedefaultseppunct}\relax
\EndOfBibitem
\bibitem[Botti \latin{et~al.}(2005)Botti, Fourreau, Nguyen, Renault, Sottile,
  and Reining]{Botti_2005}
Botti,~S.; Fourreau,~A.; Nguyen,~F. m.~c.; Renault,~Y.-O.; Sottile,~F.;
  Reining,~L. Energy dependence of the exchange-correlation kernel of
  time-dependent density functional theory: A simple model for solids.
  \emph{Phys. Rev. B} \textbf{2005}, \emph{72}, 125203\relax
\mciteBstWouldAddEndPuncttrue
\mciteSetBstMidEndSepPunct{\mcitedefaultmidpunct}
{\mcitedefaultendpunct}{\mcitedefaultseppunct}\relax
\EndOfBibitem
\bibitem[Minnhagen(1974)]{Minnhagen1974}
Minnhagen,~P. Vertex correction calculations for an electron gas. \emph{Journal
  of Physics C: Solid State Physics} \textbf{1974}, \emph{7}, 3013\relax
\mciteBstWouldAddEndPuncttrue
\mciteSetBstMidEndSepPunct{\mcitedefaultmidpunct}
{\mcitedefaultendpunct}{\mcitedefaultseppunct}\relax
\EndOfBibitem
\bibitem[Lewis and Berkelbach(2019)Lewis, and Berkelbach]{Lewis2019}
Lewis,~A.~M.; Berkelbach,~T.~C. Vertex Corrections to the Polarizability Do Not
  Improve the GW Approximation for the Ionization Potential of Molecules.
  \emph{Journal of Chemical Theory and Computation} \textbf{2019}, \emph{15},
  2925--2932\relax
\mciteBstWouldAddEndPuncttrue
\mciteSetBstMidEndSepPunct{\mcitedefaultmidpunct}
{\mcitedefaultendpunct}{\mcitedefaultseppunct}\relax
\EndOfBibitem
\bibitem[Perdew and Norman(1982)Perdew, and Norman]{Perdew1982a}
Perdew,~J.~P.; Norman,~M.~R. Electron removal energies in Kohn-Sham
  density-functional theory. \emph{Phys. Rev. B} \textbf{1982}, \emph{26},
  5445--5450\relax
\mciteBstWouldAddEndPuncttrue
\mciteSetBstMidEndSepPunct{\mcitedefaultmidpunct}
{\mcitedefaultendpunct}{\mcitedefaultseppunct}\relax
\EndOfBibitem
\bibitem[Bezanson \latin{et~al.}(2017)Bezanson, Edelman, Karpinski, and
  Shah]{Julia2017}
Bezanson,~J.; Edelman,~A.; Karpinski,~S.; Shah,~V.~B. Julia: A fresh approach
  to numerical computing. \emph{SIAM {R}eview} \textbf{2017}, \emph{59},
  65--98\relax
\mciteBstWouldAddEndPuncttrue
\mciteSetBstMidEndSepPunct{\mcitedefaultmidpunct}
{\mcitedefaultendpunct}{\mcitedefaultseppunct}\relax
\EndOfBibitem
\bibitem[Laurie(1997)]{Laurie1997}
Laurie,~D.~P. Calculation of Gauss-Kronrod Quadrature Rules. \emph{Mathematics
  of Computation} \textbf{1997}, \emph{66}, 1133--1145\relax
\mciteBstWouldAddEndPuncttrue
\mciteSetBstMidEndSepPunct{\mcitedefaultmidpunct}
{\mcitedefaultendpunct}{\mcitedefaultseppunct}\relax
\EndOfBibitem
\bibitem[Christ \latin{et~al.}(2023)Christ, Schwabeneder, Rackauckas,
  Borregaard, and Breloff]{PlotsJL}
Christ,~S.; Schwabeneder,~D.; Rackauckas,~C.; Borregaard,~M.~K.; Breloff,~T.
  Plots.jl -- a user extendable plotting API for the julia programming
  language. \textbf{2023}, \emph{11}, 5\relax
\mciteBstWouldAddEndPuncttrue
\mciteSetBstMidEndSepPunct{\mcitedefaultmidpunct}
{\mcitedefaultendpunct}{\mcitedefaultseppunct}\relax
\EndOfBibitem
\end{mcitethebibliography}
